\documentclass[aps,pra, superscriptaddress, amsmath,amssymb,nofootinbib,preprintnumber,twocolumn,showpacs]{revtex4}
\usepackage{epsfig}
\usepackage{bbm}
\usepackage{color}
\usepackage{mathrsfs,graphicx,bm,amsmath}

\newcommand{\Tr}{\mathrm{Tr}}

\newcommand{\bq}{\textbf{q}}

\begin{document}

\title{Quantum Field Theory for the Three-Body Constrained Lattice Bose Gas \\
\ Part I: Formal Developments}

\author{S. Diehl}
\affiliation{Institute for Quantum Optics and Quantum Information of the Austrian Academy of Sciences, A-6020 Innsbruck, Austria}
\affiliation{Institute for Theoretical Physics, University of Innsbruck, A-6020 Innsbruck, Austria}
\author{M. A. Baranov}
\affiliation{Institute for Quantum Optics and Quantum Information of the Austrian Academy of Sciences, A-6020 Innsbruck, Austria}
\affiliation{Institute for Theoretical Physics, University of Innsbruck, A-6020 Innsbruck, Austria}
\affiliation{RRC ``Kurchatov Institute'', Kurchatov Square 1, 123182 Moscow, Russia}
\author{A. J. Daley}
\affiliation{Institute for Quantum Optics and Quantum Information of the Austrian Academy of Sciences, A-6020 Innsbruck, Austria}
\affiliation{Institute for Theoretical Physics, University of Innsbruck, A-6020 Innsbruck, Austria}
\author{P. Zoller}
\affiliation{Institute for Quantum Optics and Quantum Information of the Austrian Academy of Sciences, A-6020 Innsbruck, Austria}
\affiliation{Institute for Theoretical Physics, University of Innsbruck, A-6020 Innsbruck, Austria}


\begin{abstract}
We develop a quantum field theoretical framework to analytically study the three-body constrained Bose-Hubbard model beyond mean field and non-interacting spin wave approximations. It is based on an exact mapping of the constrained model to a theory with two coupled bosonic degrees of freedom with polynomial interactions, which have a natural interpretation as single particles and two-particle states. The procedure can be seen as  a proper quantization of the Gutzwiller mean field theory. The theory is conveniently evaluated in the framework of the quantum effective action, for which the usual symmetry principles are now supplemented with a ``constraint principle'' operative on short distances. We test the theory via investigation of scattering properties of few particles in the limit of vanishing density, and we address the complementary problem in  the limit of maximum filling, where the low lying excitations are holes and di-holes on top of the constraint induced insulator. This is the first of a sequence of two papers. The application of the formalism to the many-body problem, which can be realized with atoms in optical lattices with strong three-body loss, is performed in a related work \cite{Diehl09II}.  
\end{abstract}

\pacs{03.70.+k,11.15.Me,67.85.-d,67.85.Hj}

\maketitle


\section{Introduction}

Lattice theories with constrained bosons have proven to be a powerful description of various spin models and strongly correlated systems in condensed matter physics \cite{AuerbachBook}. On the other hand, such theories with constrained lattice bosons have recently arisen naturally in effective models for experiments with cold atoms in optical lattices. In the presence of large two-body and three-body loss processes, bosons in an optical lattice are described on short timescales by a model with two-body and three-body constraints respectively \cite{Syassen08,Daley09}. The behavior of the Bose gas is changed drastically;  for example, in the case of the three-body constraint, the  creation of an attractive Bose gas with atomic (ASF) and dimer superfluid (DSF) phases is possible \cite{Daley09}. While the possibility of such an ASF-DSF transition has been predicted earlier in the context of continuum attractive Bose gases near Feshbach resonances \cite{Sachdev04,Radzihovsky04}, the constrained lattice system offers an intrinsic stabilization mechanism to observe such a phenomenology in experiments. This serves as one motivation to study such models theoretically in more detail, in particular exploring the consequences of the presence of the constraint. We also note that by the same dissipative blockade mechanism, constrained models with fermions may be created \cite{Han09,Kantian09}.

Here our goal is to describe the physics of a constrained boson system beyond a mean-field plus spin wave approach (see e.g. \cite{AuerbachBook}). In order to do this, we find an exact mapping of the original constrained bosonic Hubbard model to a theory of two coupled unconstrained bosonic degrees of freedom which interact polynomially. The resulting theory is conveniently analyzed in the framework of the quantum effective action, which makes it possible to study both thermodynamical and dynamical properties of the system via various many-body techniques. 
As a consequence, we can demonstrate several remarkable features of the three-body constrained attractive Bose lattice gas, which are uniquely tied to this constraint and not treated properly within a simple mean field plus spin wave approach. In particular, we show the emergence of
an Ising quantum critical point on the phase transition line between atomic to dimer superfluid phases, which generically is preempted by
the Coleman-Weinberg mechanism \cite{Coleman73} where quantum fluctuations drive the phase transition first order
, rendering the correlation length
finite \cite{Vojta00,Radzihovsky04,Sachdev04,Lee04,Balents97}. We also show the presence of a bicritical point \cite{FisherNelson74} in the strongly correlated regime at unit filling of bosons, which is characterized by energetically degenerate orders. In our case this corresponds to the coexistence of superfluidity and a charge density wave. Furthermore, quantitative effects of quantum fluctuations on the position of the phase boundary can be investigated systematically. 

The formalism we develop here has much broader applications than the bosonic lattice gas with a three-body constraint. In particular, it could be used to treat systems with effective constraints arising from large interaction parameters, and it is also applicable to constrained fermionic models, which could arise due to strong loss features in 3-component fermion gases \cite{Jochim08, Kantian09}. As a result, we focus in this paper on presenting the quantum field theoretical construction in detail, and give benchmark calculations for this method. Its application to the many-body attractive lattice Bose gas with three-body constraint is left to \cite{Diehl09II}, where we discuss in detail the results outlined above. The results of the present and the related paper \cite{Diehl09II} are summarized in \cite{Diehl09Short}, where we also indicate how to probe our findings in experiments with ultracold atoms.

This paper is organized as follows. We begin by reviewing the microscopic derivation of the three-body constrained model via a dissipative blockade mechanism in Sec. \ref{sec:MicDeriv}. In Sec.~\ref{sec:overview} we give an overview of our formalism and a comparison to existing methods for treating constrained models. Sec. \ref{QFTBH} contains the central result of this paper, the mapping of the constrained model to a coupled boson theory already anticipated above. In Sec. \ref{VacProb} we apply the formalism to the two limits where true many-body effects are absent, $n=0$ and maximum filling $n=2$. There our boson model reduces to Feshbach-type models, which we analyze in terms of Dyson-Schwinger equations \cite{Dyson49Schwinger51}. We perform the above mentioned benchmark calculations at $n=0$. Furthermore, we investigate hole/di-hole scattering and bound state formation at $n=2$, and present a fourth order perturbative calculation for the dimer-dimer interaction. This will be needed as an input for the many-body theory in \cite{Diehl09II}. Our conclusions are drawn in Sec. \ref{sec:Conclusion}.

\section{Derivation of the Constrained Microscopic Model}
\label{sec:MicDeriv}

In this section, we review how the three-body hardcore constraint for the Bose-Hubbard model emerges from strong three-body loss \cite{Daley09}. 

We consider bosons in the lowest Bloch band of an optical lattice, which are described by the Bose-Hubbard Hamiltonian ($\hbar\equiv 1$), 
\begin{eqnarray}
H_{\rm BH} &=&-J\sum_{\langle i,j\rangle}a_{i}^{\dagger}a_{j} - \mu \sum_{i}a_{i}^{\dagger }a_{i}
+\tfrac{1}{2}U\sum_{i}a_{i}^{\dagger \,\,2}a_{i}^{2},
\end{eqnarray}
where $a_i (a_i^\dag)$ is the bosonic annihilation (creation) operator at site $i$. $J$ is the hopping rate, $\mu$ the chemical potential, and $U$ the onsite interaction. The convention $\langle i,j\rangle$ first sums over all sites $i$, and then over the neighbourhood of each $i$ spanned by sites $j$, $\sum_{\langle i,j\rangle} = \sum_i  \sum_{\langle j| i\rangle}$. This model is valid in the limit where $J, Un \ll \omega$, where $\omega$ is the separation between Bloch bands and $n$ is the mean density.

Three-body loss in this system is due to inelastic collisions of three atoms, two of which form a deeply bound molecule. Together with the third atom, molecule formation is compatible with energy and momentum conservation, unlike the case of two-particle collisions. Since the binding energy of the molecule typically strongly exceeds optical lattice depths, the resulting kinetic energy of the products couples them to the continuum of unbound states, thus leading to their escape from the lattice. This picture allows us to write down a zero temperature master equation in the Markov approximation, which in the simplest approximation neglects loss arising from particles on neighbouring lattice sites. This master equation is given by
\begin{eqnarray}\label{Diss1}
\dot \rho &=& -\mathrm i\left(H_{\rm eff} \rho - \rho H_{\rm eff}^\dag \right) +\frac{\gamma_3}{12} \sum_i 2  a_i^3 \rho a_i^{\dag \,3},\\\nonumber
H_{\rm eff}&=& H-\mathrm i \frac{\gamma_3}{12} \sum_i a_i^{\dag \,3} a_i^3,
\end{eqnarray}
where the decay rate $\gamma_3$ can be roughly estimated from the experimentally measured continuum loss rate via the usual Wannier construction \cite{Daley09}. We have absorbed the Schr\"odinger type terms of the dissipative evolution into an effective Hamiltonian with imaginary, and thus decay, term. The remaining recyling term couples sectors in the density matrix with particle number $N,N-3, ...$ and ensures a norm conserving time evolution. 

We are now interested in the limit of strong loss, $\gamma_3 \gg J, U$, which suggests a perturbative expansion in $1/\gamma_3$. The scale $\gamma_3$ only couples states with three and more particles per site. We therefore define a projector $P$ onto the subspace with at most 2 atoms per site, and its complement as $Q=1-P$. To second order in perturbation theory we then find the projected Hamiltonian 
\begin{eqnarray}\label{Diss2}
H_{P, {\rm eff}} &\approx& P H_{\rm BH} P + P H_{\rm BH} Q ( Q H_{\rm BH} Q)^{-1} Q H_{\rm BH} P\nonumber\\
& =& P H_{\rm BH} P - \frac{\mathrm i \Gamma}{2} \sum_i P c_i^{\dag} c_i P, 
\end{eqnarray}
and the corresponding Master Equation is
\begin{eqnarray}
\dot \rho_P = -\mathrm i\left(H_{P,\rm eff} \rho_P - \rho_P H_{P,\rm eff}^\dag \right) +\Gamma \sum_i P c_i \rho_P c_i^{\dag \,3}P,
\end{eqnarray}
with $\rho_P=P\rho P$. The effective decay rate and jump operators are given by 
\begin{eqnarray}
\Gamma &\approx& 12\frac{J^2}{\gamma_3}, \quad c_i =  a_i^2 \sum_{\langle j|i\rangle} a_j \,\, /\sqrt{2}.
\end{eqnarray}
The respective terms in $H_{P,\rm eff}$ have simple interpretations: The leading term describes the coherent dynamics of lattice bosons, but with the constraint of not populating a single site with more than two particles. This projected part $H =  P H_{\rm BH} P$ can thus be written as a 3-body constrained Bose-Hubbard Hamiltonian
\begin{eqnarray}\label{ConstraHam}
H &=& -J\sum_{\langle i,j\rangle}a_{i}^{\dagger}a_{j} - \mu \sum_{i}a_{i}^{\dagger }a_{i}+\tfrac{1}{2}U\sum_{i}a_{i}^{\dagger \,\,2}a_{i}^{2},\nonumber\\ a_i^{\dag \, 3} &\equiv& 0.
\end{eqnarray}
The leading correction is imaginary and describes particle number loss. The decay rate $\Gamma\sim 1/\gamma_3$ is, however, very small in the considered limit. Consequently, over timescales $\tau =1/\Gamma$, one realizes indeed the physics of the Bose-Hubbard model with 3-body hardcore constraint. In \cite{Daley09}, we have shown that e.g. in atomic Cesium systems close to the zero crossing of the scattering length, the loss rate $\gamma_3$ is the dominant energy scale. There, we have also analyzed the many-body dissipative dynamics of Eq. \eqref{Diss1} in the regime described by Eq. \eqref{Diss2} with exact DMRG methods in one spatial dimension, including the specification of a scheme with which the ground state of $PHP$ can be reached. This motivates a more detailed analytical investigation of the zero temperature phase diagram of Eq. \eqref{ConstraHam}. As mentioned in the introduction, such a scenario with dominant three-body loss also arises naturally in three-component fermion systems close to a Feshbach resonances \cite{Jochim08}, where it arises due to the proximity to rapidly decaying Efimov state.

\section{Overview}
\label{sec:overview}

In this section, we provide an overview and discussion of our construction and compare it with existing theoretical approaches. 

The starting point is a truncation of the onsite bosonic Hilbert space to three states corresponding to zero, single and double occupancy. Following Altman and Auerbach \cite{Altman02}, we introduce three operators creating these states out of an auxiliary vacuum state. The operators are not independent but obey a holonomic constraint, such that one of the degrees of freedom can be eliminated. We propose a resolution of this constraint in a fashion that (i) produces polynomial interactions between the two remaining operators and (ii) allows us to interpret them as standard bosonic degrees of freedom. The Hamiltonian written in terms of these bosonic operators is an involution on the physical Hilbert space (associated to onsite occupation $\leq 2$), i.e. all matrix elements coupling into the unphysical sector with onsite occupation $>2$ vanish. Consequently the partition sum for this Hamiltonian decomposes into a physical and an unphysical part. In order to distinguish the respective contributions, we choose to calculate the quantum effective action from a functional integral representation of the partition sum. This is the most general polynomial in the operators of the theory which is compatible with the symmetries of the microscopic Hamiltonian, respectively the microscopic action. We argue that the general restrictions imposed by symmetry are supplemented by a further new ``constraint principle'', the requirement that the contributions to this polynomial be compatible with the constraints imposed by the microscopic Hamiltonian -- the evaluation can proceed as in a standard polynomial boson theory. This opens up the powerful toolbox of modern quantum field theoretical methods for calculations in onsite constrained models, and the effective action provides a unified framework to treat the strong correlations at short distances characteristic to lattice superfluids as well as the deep infrared sector of the theory, where questions on the nature of the phase transition can be addressed. Similar to symmetries, the restrictions imposed by the constraint on the full theory leverage over from the microscopic theory. However, unlike symmetries, the relevance of the constraint depends on scale, being restrictive on short distances, while on long distances power counting arguments lead to an effectively unconstrained though interacting spin wave theory.


Our approach may be conceived as a proper quantization of the Gutzwiller mean field theory: The construction implies that the zero order contribution to the thermodynamic effective potential is the Gutzwiller mean field energy. The quadratic fluctuations reproduce spin wave theory. However, due to the exact nature of the mapping we can also assess the effects of interactions systematically. At the same time, this shows that while the mapping is exact, it builds on the knowledge of qualitative properties of the ground state of the system, such as the symmetry breaking pattern. In other words, given this pattern, we can systematically construct the excitations and their interactions on top of it. The bias introduced by the assumptions on the ground state is actually shared with all analytical approaches to many-body systems involving symmetry breaking. 


One established way to deal with constrained bosonic theories with a finite number of onsite states would be to map it to the corresponding spin model (spin 1 in our case), and subsequently analyze this model. In practice, however, the analytical evaluation of the spin model beyond a free spin wave level is technically challenging. Furthermore, the spin language is not particularly advantageous and physically intuitive for models which originate from a constrained boson theory, cf. Sect \ref{sec:spin1}. We see a central advantage of our formalism vs. such a spin model approach in the straightforward interpretation of the two bosonic fields in physical terms, which are close to the original degrees of freedom. For example, at low density $n\gtrsim 0$ and close to maximal filling $n\lesssim 2$ they may be interpreted as atoms and bound states of these, dimers, resp. holes and di-holes, which are indeed expected to form the dominant low energy excitations in the respective situations. At intermediate densities hybridization takes place. The microscopic Hamiltonian reduces to Feshbach-type models, with the characteristic splitting term of dimers/di-holes into atoms. The role of the detuning is now played by the interaction term $U$. We may view this result as a built-in Hubbard-Stratonovich transformation on the level of the Hamiltonian, which importantly respects the constraint. At this point, we note an important difference of our approach to the one by Dyson and Maleev \cite{DysonMaleev}, which map a spin model to a model with a single bosonic degree of freedom. The generalization of a Hubbard-Stratonovich transformation to such a model would be problematic. 

Other related approaches to constrained models have been put forward in the context of slave boson theories for the description of strongly correlated fermions \cite{BarnesColeman,KotliarAnderson}, or in the Schwinger-boson or Holstein-Primakoff approach to spin models \cite{AuerbachBook}. While these formulations are in principle exact, the practical implementation of the constraint is done on an approximate level only \cite{KotliarAnderson}, or in a fashion that makes it difficult to assess coupling constants of an effective low energy theory or thermodynamic quantities  \cite{BarnesColeman}. Our approach is tailored to make these accessible. Furthermore, we stress that our exact implementation of the constraint differs from the conventional approximate treatment via expansion of the square root as done e.g. for the case of large spin $S$ in the Schwinger-boson or Holstein-Primakoff approaches \cite{AuerbachBook}. We provide concrete benchmarks for our procedure through the analysis of the vacuum problem of a few scattering particles. While our approach yields the correct non-perturbative Schr\"odinger  equation for two-particle scattering and the right coefficient for an induced nearest-neighbour dimer-dimer interaction, the square root expansion would produce wrong prefactors.

Clearly, going beyond the mean field approach in constrained bosonic models is nowadays possible making use of powerful numerical techniques such as Quantum Monte Carlo \cite{Kohno97,Batrouni99,Bernadet02} or variational simulations \cite{Murg07}. However, often a more analytical understanding of the physics leading to a certain quantitative effect is desirable. Furthermore, as we have mentioned above, our formalism can be extended to fermion systems, which is not straightforward with numerically exact techniques.

\section{Mapping the Constrained Model to an Interacting Boson Theory}
\label{QFTBH}

In this section we derive the mapping of the bosonic model with three-body hardcore constraint Eq.  \eqref{ConstraHam} to an unconstrained model with two bosonic degrees of freedom, coupled via polynomial interactions. This construction depends on the filling of the lattice. We concentrate on the ``vacuum limit'' of zero density and temperature first, where one deals with a few  particles and many-body effects, such as spontaneous symmetry breaking, are absent. The generalization to arbitrary filling $0\leq n \leq 2$ is performed in Sec. \ref{sec:RotNew}.

\subsection{Reduction to three on-site states}

We start from the 3-body constrained Bose-Hubbard Hamiltonian Eq. \eqref{ConstraHam}. The constraint makes it possible to restrict the Hilbert space on each site to three states with occupation 0,1,2. We introduce three operators that create "particles" in these states. Such a slave-boson type procedure has been proposed previously by Altman \emph{et al.} \cite{Altman02} in the context of the repulsive Bose-Hubbard model, where it constitutes an approximation. Here this step is exact:
\begin{eqnarray}\label{ConstraintOps}
|\alpha \rangle &=& t_{\alpha,i}^\dag |\text{vac}\rangle = \frac{1}{\sqrt{\alpha!}} \,\, a_i^{\dag \,\, \alpha} |\text{vac}\rangle, \quad \alpha =0,1,2.
\end{eqnarray}
The operators $t_{\alpha, i}$ are so far only defined by their action of the ``vacuum'' state $|\text{vac}\rangle$ and we discuss their commuation relations below. They obey a holonomic constraint,
\begin{eqnarray}\label{HOL}
 \sum_{\alpha =0}^2 t^\dag_{\alpha, i}t_{\alpha, i} = \textbf{1}.
\end{eqnarray}
In the space spanned by $|0\rangle,|1\rangle,|2\rangle$ we may express the creation operator as
\begin{eqnarray}\label{Inversion}
a_i^\dag =  \sqrt{2}  t_{2,i}^\dag t_{1,i} + t_{1,i}^\dag t_{0, i}.
\end{eqnarray}
Hence we can express the original Hamiltonian equivalently in terms of the new operators, defining non-hermitian kinetic operators 
\begin{eqnarray}
K^{(10)}_i = t_{1,i}^\dag  t_{0,i},\quad K^{(21)}_i = t_{2,i}^\dag  t_{1,i},
\end{eqnarray}
and hermitian potential energy operators
\begin{eqnarray}
\hat n_{1,i} = t_{1,i}^\dag  t_{1,i},\quad \hat n_{2,i} = t_{2,i}^\dag  t_{2,i},
\end{eqnarray}
we obtain
\begin{eqnarray}\label{Ht}
H_{\text{kin}} &=& - J \sum_{\langle i,j\rangle}\big[ K^{(10)}_i K^{(10)\dag}_j+  2 K^{(21)\dag}_i K^{(21)}_j \\\nonumber 
&&\qquad \quad + \sqrt{2} (K^{(21)}_i K^{(10)\dag}_j +K^{(10)}_i K^{(21)\dag}_j) \big], \\\nonumber
H_{\text{pot}} &=& -\mu \sum_i  \hat n_i + U\sum_{i} \hat n_{2,i},\quad \hat n_{i}  =   \hat n_{1,i}  + 2 \hat n_{2,i}.
\end{eqnarray}
We observe that the local terms become simple (quadratic) in this representation, while the nonlocal terms are fourth order in the operators, giving rise to ``kinematic'' interactions. The onsite interaction part therefore can be treated exactly, while the complexity of the problem is now encoded in the hopping term. This is reminiscent of the conventional Gutzwiller (mean field) approach to the Bose-Hubbard model. Indeed, implementing a Gross-Pitaevski-type mean field theory for the above Hamiltonian by formally replacing the operators by complex valued amplitudes $t_{\alpha,i} \to f_{\alpha,i}$, the above Hamiltonian operator reduces to the Gutzwiller energy expression $E_{\text{GW}} = \langle\psi | H |\psi\rangle$ for the case that the wave function $ |\psi\rangle= \prod_i  |\psi\rangle_i$ is truncated to the three lowest Fock states on each site, $ |\psi\rangle_i = \sum_{\alpha=0}^{2} f_{\alpha,i} |\alpha\rangle_i$. In this case, the holonomic constraint reduces to the normalization condition for the wave function, $_i\langle\psi |\psi\rangle_i =\sum_{\alpha} f^*_{\alpha, i}f_{\alpha, i} = 1 \,\forall\, i$. 

It may be tempting to try to develop a many-body theory including the description of spontaneous symmetry breaking at finite density for the above Hamiltonian directly in terms of the $t_\alpha$ operators in the sense of a Bogoliubov-type theory on top of the Gross-Pitaevski mean field, via a replacement of the type $t_{\alpha,i}  = f_{\alpha,i} + \delta t_{\alpha,i}$ familiar from low density continuum theories. Such a procedure, however, leads to severe consistency problems when encompassing the full range of densities allowed by the three-body hardcore constraint, $0\leq n \leq 2$. Therefore, we first focus on the ``vacuum limit'' $n=0$ where no spontaneous symmetry breaking is present, and where the Hamiltonian \eqref{Ht} describes the physics of a few scattering particles. The generalization to arbitrary density is performed in Sec. \ref{sec:RotNew}. 

\subsection{Implementation of the Holonomic Constraint: Interacting Boson Theory}
\label{HolConstraint}

The holonomic constraint is now used to eliminate one of the operators. In the limit $n=0$, all the amplitude resides in the zero-fold occupied state, and the mean field vacuum is described by $|\Omega \rangle = \prod_i t_{0,i}^\dag |\text{vac}\rangle$. We thus eliminate the operators $t_{0,i}$. The remaining two operators describe excitations on top of this mean field vacuum, and will have a natural interpretation in terms of atoms and dimers. At this point, it is transparent that our construction builds on the proper choice of the \emph{qualitative} features of the physical vacuum. This prerequisites a certain understanding of the physics, and introduces a bias in our construction. The subsequent construction is however exact, and can be used to \emph{quantitatively} calculate properties of the system. The situation is actually similar to the treatment of interacting Bose gases in the continuum, where a condensation of the particles in the zero mode is assumed to expand in the fluctuations around this mean field. 

The holonomic constraint can, in principle, be implemented by the replacements
\begin{eqnarray}\label{NaiveConstraint}
t_{1,i}^{\dag}t_{0,i}&\rightarrow& t_{1,i}^{\dag}\mathrm e^{\mathrm i \varphi_i}\sqrt{1- \hat n_{1,i} - \hat n_{2,i}} , \\\nonumber
 t_{0,i}^{\dag}t_{1,i}&\rightarrow&\sqrt{1- \hat n_{1,i} - \hat n_{2,i}}\,\,\mathrm e^{-\mathrm i \varphi_i} t_{1,i}.%
\end{eqnarray}
The holonomic constraint is local and only restricts the amplitude of $t_{0,i}$, therefore we introduce a phase-amplitude representation $|t_{0,i}|\exp \mathrm i \varphi_i$. We first discuss the role of this phase and then turn to the more interesting question of the amplitude. 

Inserting the replacement into Eq. \eqref{Ht}, we observe that a \emph{local} redefinition of $t_1$ and $t_2$ as 
\begin{eqnarray}\label{RotFrame}
t_{1,i}\to \tilde t_{1,i}= t_{1,i}\mathrm e^{\mathrm i \varphi_i},\quad  t_{2,i}\to \tilde t_{2,i}= t_{2,i}\mathrm e^{\mathrm i \varphi_i}
\end{eqnarray}
for all $i$ completely removes the phase from the Hamiltonian. Therefore, we can work in this rotated frame from the outset, and simply consider the two complex valued operators $ \tilde t_{1;2,i}$. We will drop the tilde in the following. 

Now we study the amplitude. Obviously the square roots are impracticable for any field theory calculation where one has to work with polynomials in the field operators. However, on our subspace, the matrix elements of both $(1- \hat n_{1,j} - \hat n_{2,j})^{1/2}$ and $(1- \hat n_{1,j} - \hat n_{2,j})$ are the same: either $1$ or $0$. Consequently, on the subspace we may replace 
\begin{eqnarray}\label{Constraint}
t_{1,i}^\dag  t_{0,i} &\to& t_{1,i}^\dag X_i , \quad t_{0,i}^\dag t_{1,i} \to X_i t_{1,i}, \\\nonumber
 X_i &=& 1- \hat n_{1,i} - \hat n_{2,i}.
\end{eqnarray}
Note that the second expression could also be replaced by $(1- \hat n_{1,i} - \hat n_{2,i})^{1/2}t_{1,i}\rightarrow t_{1,i}$, since the action of this operator is nonzero only if originally there is an atom on the $i$-th site, but when this atom is annihilated by $t_{1,i}$ with an empty state left, the action of the square root is simply unity. However, for hermeticity issues we prefer to work with the variant in Eq. \eqref{Constraint}.

Formally, we may justify the replacement of the square root by a polynomial by the following formula: Consider a linear operator $\tilde X$  with the property $\tilde X^2=\tilde X$. Then for a function $f$ of this operator one has, using the Taylor representation,
\begin{eqnarray}
f(\tilde X) &=& \sum\limits_{n=0}\limits^\infty \frac{f^{(n)}(0)}{n!}\tilde X^n = f(0)  + \tilde X \sum\limits_{n=1}\limits^\infty \frac{f^{(n)}(0)}{n!}\,\, 1 \nonumber\\
&=& f(0) (1 - \tilde X) + \tilde X \sum\limits_{n=0}\limits^\infty \frac{f^{(n)}(0)}{n!}\,\,1\nonumber\\
&=& f(0) (1-\tilde X)+ f(1)\tilde X.
\end{eqnarray}
In our case, we have $\tilde X = 1 - X_i = \hat n_{1,i} + \hat n_{2,i}$ for all $i$, and $f(\tilde X) = \sqrt{ 1 - \tilde X}$. Indeed, $\tilde X^2 =\tilde X$. Seen as a function of $X_i = 1 - \tilde X$, we have $f(X_i) = \sqrt{X_i} =  f(0) X_i+ f(1)(1- X_i)$. Since also $X_i^2= X_i$ the latter result would have been obtained from the Taylor representation of $f(X_i)$ around $0$.  The auxiliary operator $\tilde X$ is introduced to circumvent an expansion of the square root around $0$, but leads to the same result. No approximation has been used here. 

Having implemented the constraint, we are now going to show that the remaining operators $t_1,t_2$ can be treated as \emph{standard bosonic operators}. Consequently the Hamiltonian (or the corresponding action) with the above replacements will lend itself for a treatment with well established field theoretic methods. To show that we may interpret $t_1,t_2$ as bosonic operators, we assume a bosonic Hilbert space for atoms $t_1$ and dimers $t_2$ at each site $i$, which reads $\mathcal H_i = \{ | n_i\rangle  |m_i\rangle \}, n_i, m_i = 0,1, ...$. The complete Hilbert space is $\mathcal H =\prod_i \mathcal H_i$. We divide the onsite Hilbert spaces into a physical subspace $\mathcal P_i$ and an unphysical one $\mathcal U_i$, $\mathcal H_i  = \mathcal P_i \oplus \mathcal U_i$; the subspaces are orthogonal by construction. The physical subspace is spanned by the combinations 
\begin{eqnarray}
\mathcal P_i = \{ | 0_i\rangle  |0_i\rangle , | 1_i\rangle  |0_i\rangle , | 0_i\rangle  |1_i\rangle \}.
\end{eqnarray}
In our construction of introducing atom and dimer operators, the state with two atoms on one site is represented as one dimer $| 0_i\rangle  |1_i\rangle$; $| 2_i\rangle  |0_i\rangle$ instead is already part of $\mathcal U_i$. 

A first observation is that standard bosonic operators have the same action on the physical subspace as the original operators $t_1,t_2$ defined via Eq. (\ref{ConstraintOps}), since the bosonic $\sqrt{n}$-enhancement ($b^\dag |n\rangle =\sqrt{n+1} |n+1\rangle$ for bosonic operators $b^\dag$) is either 0 or 1 on the physical subspace. The assumption of $t_1,t_2$ being bosons is thus consistent on the physical subspace.

Next we consider the Hamiltonian Eq. (\ref{Ht}) with the constraint implemented via Eq. \eqref{Constraint}, with the goal to show that   the time evolution generated by this Hamiltonian \emph{does not couple} the two subspaces. It reads
\begin{eqnarray}\label{HC}
H_{\text{kin}} &\equiv& \hspace{-0.2cm} \sum_{\langle i,j \rangle } H_{i,j} \hspace{-0.05cm}=\hspace{-0.05cm}  - J \hspace{-0.08cm}\sum_{\langle i,j\rangle}\big[ t_{1,i}^\dag X_i X_j t_{1,j} 
+ 2 t_{2,i}^\dag t_{2,j}t_{1,j}^\dag t_{1,i} \nonumber\\
&& \qquad\quad \qquad  + \sqrt{2} (t_{2,i}^\dag  t_{1,i}X_j t_{1,j} + t_{1,i}^\dag X_i t_{1,j}^\dag t_{2,j}  ) \big] ,\nonumber\\
H_{\text{pot}} &=&  \sum_i (U - 2\mu)  \hat n_{2,i} -\mu \hat n_{1,i},
\end{eqnarray}
In the kinetic term, the first expression describes the conditional hopping of single atoms, the second represents the exchange of a dimer and an atom on neighbouring sites, and the last one describes the conditional bilocal splitting and recombination of a dimer into atoms. It can be easily shown that $H$ maps physical on physical states, and unphysical on unphysical ones, while there are no transitions between the subspaces generated by \eqref{HC}. For that purpose it is sufficient to check that 
\begin{eqnarray}
H_{i,j} | n_i\rangle |m_i\rangle |n_j \rangle | m_j\rangle 
\end{eqnarray}
is in $\mathcal P_i\mathcal P_j$ if the initial state is in $\mathcal P_i \mathcal P_j$ (a simple 9 dimensional space). Since this means that all matrix elements $\langle u | H_{ij} | p\rangle = 0$, this is sufficient to conclude that starting in $\mathcal U$ the mapping will be into $\mathcal U$, since for the hermitian $H$ one has $\langle p | H_{ij} | u\rangle = \langle u | H_{ij} | p\rangle^* = 0$. 
In consequence $H$ can be written in the form
\begin{eqnarray}
H = H_{P} \otimes \textbf{1}_{U} + \textbf{1}_{P} \otimes H_{U} .
\end{eqnarray}
In other words, $\mathcal P = \prod_i \mathcal P_i$ and $\mathcal U= \prod_i \mathcal U_i$ are invariants under application of $H$, and thus also repeated application does not lead out of the subspaces. Therefore, we also have 
\begin{eqnarray}
\exp (- \beta H)  =  \exp (- \beta H_{P})  \otimes \textbf{1}_{U} + \textbf{1}_{P} \otimes \exp (- \beta H_{U}) .\nonumber\\
\end{eqnarray}
The partition sum is the given by 
\begin{eqnarray}
Z &=& \Tr \exp (-\beta H)  \\\nonumber
&=& \hspace{-0.2cm} \sum\limits_{\{p,u\}} ( \langle p|, \langle u | )
\left(\begin{array}{cc} 
\exp   (-\beta H_{P})  & 0 \\
0 & \exp (- \beta H_{U}) 
\end{array}\right) \left(\begin{array}{c} |p\rangle \\ |u\rangle \end{array}\right)\\\nonumber
&=&  \sum\limits_{\{p\}}  \langle p| \exp (- \beta H_{P})   |p\rangle  + \sum\limits_{\{u\}}  \langle u| \exp (- \beta H_{U})   |u\rangle .
\end{eqnarray}
Thus we get contributions from both the physical and the unphysical part of the Hilbert space. However, the key point is that they do not mix; thus the answers found for the physical part will be correct, and we only need to find the criterion to discriminate the physical from the unphysical part of the partition sum. 

This issue is addressed in the last step of the construction. Indeed, such a setting is provided by using the \emph{effective action} to encode the physical information of the theory, as we will now outline. First we represent the partition function as a Euclidean functional integral, which is straightforward as we simply have to quantize a theory with two coupled bosonic degrees of freedom \cite{NegeleOrland}:
\begin{eqnarray}\label{ActHam}
Z   &=& \int \mathcal D t_1 \mathcal D t_2 \exp - S [ t_1 , t_2], \\\nonumber
S &=& \int d \tau \Big( \sum_i t_{1,i}^\dag \partial_\tau t_{1,i} +  t_{2,i}^\dag \partial_\tau t_{2,i} + H[t_1,t_2]\Big),
\end{eqnarray}
where $H$ is the Hamiltonian above, however to be interpreted in the Heisenberg picture with (imaginary) time dependent fields, and the fields are now classical fluctuating variables. $S$ is the classical Euclidean action. In the next step we introduce a source term in the partition function,
\begin{eqnarray}
Z[j_1,j_2] &=& \int \mathcal D t_1 \mathcal D t_2 \exp - S [ t_1 , t_2] \\\nonumber
&&\qquad + \int d\tau \sum_i (j_{1,i}^\dag t_{1,i} + j_{2,i}^\dag t_{2,i} + c.c. ), \\\nonumber
Z &=& Z[j_1=j_2 =0].
\end{eqnarray}
The source terms introduce linear terms in $t_{1,2}$, which mix the physical and the unphysical sectors. Since $j_{1,2}$ is only used in a pivotal sense to generate the correlation functions upon functional differentiation, and set to zero at the end of the calculation, this does not pose any conceptual problems. The situation is analogous to the effect of the source term on the symmetries of the theory, which are broken explicitly for nonzero sources. 

The effective action is defined as the Legendre transform of the free energy $W[j] = \log Z[j]$ (we introduce the shorthands $\hat\chi = (t_1,t_1^\dag, t_2, t_2^\dag), j= (j_1,j_1^\dag, j_2, j_2^\dag)$  \cite{AmitBook}:
\begin{eqnarray}
\Gamma [\chi] = - W[j] + \int j^T \chi, \quad \chi \equiv \frac{\delta W[j] }{\delta j} ,
\end{eqnarray}
where $\chi = \langle \hat \chi\rangle$ is the field expectation value or the ``classical'' field. By the Legendre transform, the active variable is changed from $j$ to $\chi$. The effective action has the following representation in terms of a functional integral,
\begin{eqnarray}\label{EAPI}
\exp - \Gamma[\chi] &=& \int \mathcal D \delta \chi \exp - S [ \chi + \delta \chi] + \int j^T \delta \chi , \nonumber\\
 j &=& \frac{\delta \Gamma [\chi]}{\delta \chi },
\end{eqnarray}
where $\delta \chi  \equiv \hat \chi  - \chi$. The last identity is the full quantum equation of motion, and the equilibrium situation we are interested in is specified by $j=0$ where no mixing between the physical and the unphysical sector occurs. When fluctuations are unimportant, the integration over the $\delta \chi$ can be dropped and the above equation reduces to $\Gamma [\chi] = S[\chi]$, i.e. the quantum effective action reduces to the classical one.

The effective action expresses the theory in terms of the fields $\chi$. The vertex expansion generates the one-particle irreducible (1PI) correlation functions, 
\begin{eqnarray}
\Gamma [ \chi ]  = \sum_l \frac{1}{l !} \int_{x_1, ...., x_l}\hspace{-0.5cm} \Gamma_{i_1, ... , i_l}^{(l)}(x_1, ..., x_l) \chi_{i_1}(x_1) \cdot ... \cdot \chi_{i_l}(x_l).\nonumber\\\hspace{-0.5cm}
\end{eqnarray}
Usually, the coupling coefficients of the expansion are only restricted by the symmetries of the theory -- the effective action is the most general polynomial in the fields $\chi$ which is compatible with the latter. Thus, formulating the theory in terms of physical objects -- the fields $\chi$ -- offers the advantage of directly leveraging the power of symmetry considerations from the microscopic (or classical) to the full quantum level. 

In complete analogy, we can make use of the restrictions present in the microscopic Hamiltonian when computing the quantum effective action. Since, as we have shown above, no couplings mapping from $\mathcal U \to \mathcal P$ are generated, we may write down the most general form for the effective action for the physical sector of the theory by directly excluding couplings which would violate this constraint. In practice, this concerns processes which change the on-site occupation number. For example, a process which involves creation of a dimer on site $i$ must be accompanied by an appropriate constraint that the site be empty prior to the process. Thus, the operator $t_2^\dag$ must always appear in the combination $t_2^\dag X_i$. 
Furthermore requiring hermeticity of the terms appearing in the effective action, we conclude that the effective nearest-neighbour dimer hopping term is of the form $J_{\text{eff}}t^\dag_{2,i}X_i X_jt_{2, j} $. In the practical calculation, we may restrict ourselves to the computation of the coefficient which is simplest to extract -- obviously the quadratic one.  

In sum, we have obtained the following result: the usual symmetry constraints on the quantum effective action are now supplemented by a further fundamental principle, namely the restrictions present in the microscopic theory which originate from the hardcore constraint. 

\subsection{Arbitrary density}
\label{sec:RotNew}

The construction presented in the last section focused on the zero density limit, describing the scattering of few particles in the absence of many-body effects. At a finite density, the low temperature physics of bosons is characterized by the spontaneous breaking of global phase rotation symmetry $U(1)$. The ground state exhibits a condensate mean field, which has to be incorporated in the theoretical description of the system. One customary approach is to quantize a theory with degrees of freedom $\hat b = s +\delta b$ via the path integral, where $\delta b$ is the fluctuation around the classical field $s$. However, in our case this procedure does not work, since possible values of the mean field lie on the compact interval $[0,2]$, and  due to the nonlinear nature of the constraint. To this end we follow Huber \emph{et al.} \cite{Altman07}, who implement the procedure on the mean field plus spin wave level. We will see that our treatment of the constraint can be applied also in this case, such that we arrive at an exact formulation of the problem in the presence of spontaneous symmetry breaking. The procedure consists in first introducing the mean field via a unitary rotation in the space of operators (Eq. \eqref{ratation} below), and then quantizing this theory of operators which are free of expectation values via the path integral.  In this way, we can obtain a picture which is fully consistent with general features of many-body theories with $U(1)$ symmetry, in particular, we can derive Goldstone's theorem within our framework \cite{Diehl09II}. 

Here we concentrate on homogeneous ground states, noting that the implementation of spatially dependent order parameters -- such as a charge density wave -- is straightforward. One then simply has to use rotation matrices which vary from site to site. Such a situation will be encountered in \cite{Diehl09II}.

\begin{figure}
\begin{center}
\vspace{-2.7cm}
\includegraphics[width=1\columnwidth]{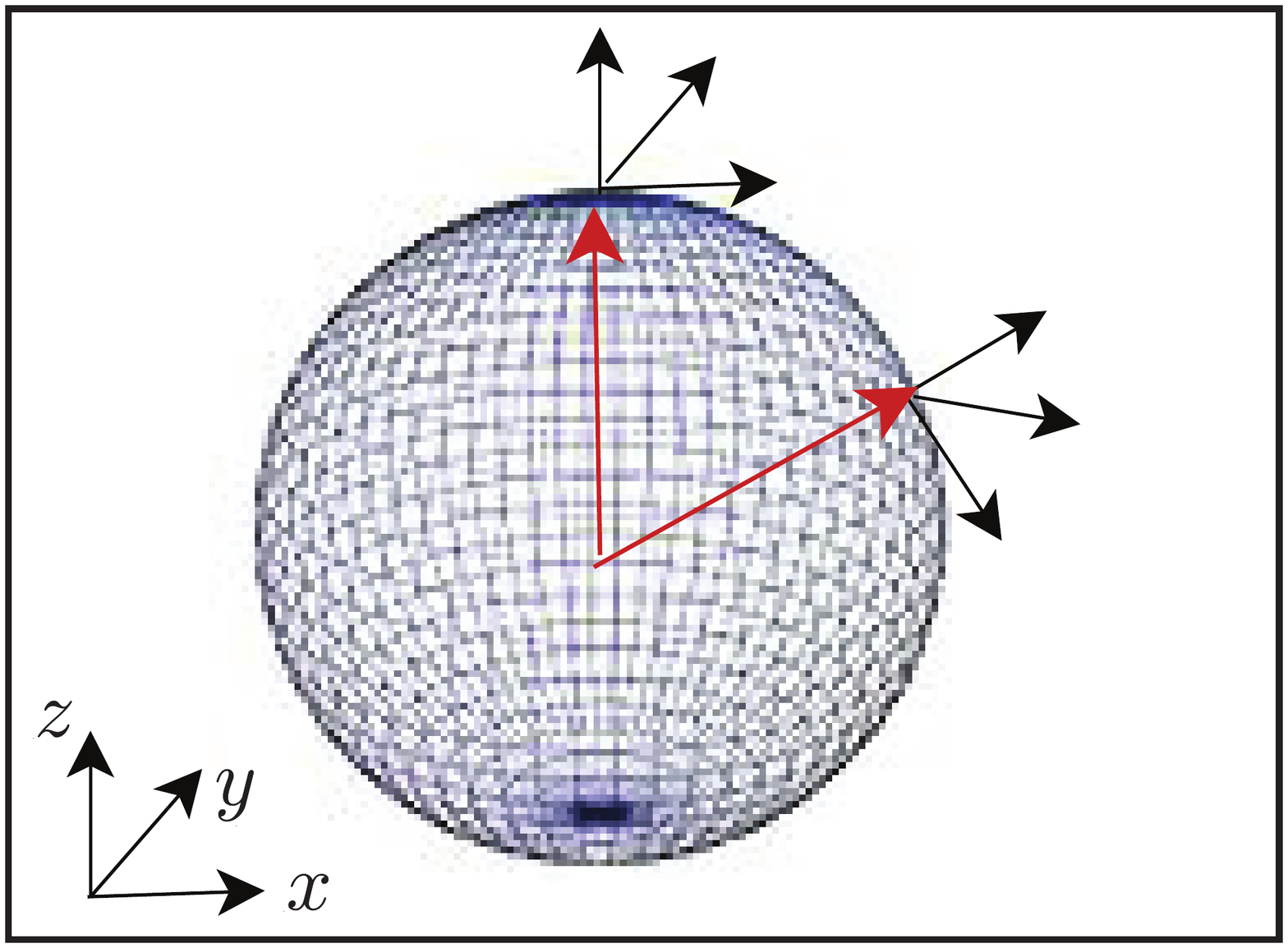}
\end{center}
\vspace{-3.4cm}
\caption{\label{Fluctuations} Rotation to a new ground state: The vacuum states $n=0 (2)$ are described by the red mean field vector pointing in positive (negative) $z$ direction. The fluctuations (black arrows) form a coordinate system in which the direction collinear to the mean field vector is eliminated via the implementation of the constraint. All mean field vectors not in the $z$ direction describe a homogeneous superfluid ground state. A situation analogous to the one for the vacua is achieved via an appropriate rotation of the coordinates for the fluctuations.}
\end{figure}

Our treatment of the vacuum problem at $n=0$ started from the idea that in this case, all the amplitude resides in the zero-fold occupied state, and that fluctuations around this state have to be considered. The excitations on top of this ``mean field'' vacuum, defined as $|\Omega\rangle = \prod_i t_{0,i}^\dag |\text{vac}\rangle$, then turned out to be single atoms and dimers, respectively, as expected intuitively, and we have formulated the corresponding quantum field theory to describe their scattering properties. 

In the many-body problem, we proceed in complete analogy by first introducing a mean field vacuum. A general homogeneous mean field vacuum may be written as \footnote{The phase relation between the amplitudes $f_\alpha = r_\alpha e^{\mathrm i \theta_\alpha}, \theta_\alpha =\alpha \phi$ is a consequence of spontaneous symmetry breaking in Fock space. $\phi$ is the spontaneously chosen phase of the condensate.}
\begin{eqnarray}\label{amplitudesI}
|\Omega \rangle &=& \prod_i \big(\sum\limits_\alpha r_\alpha \exp ( i \alpha\phi) |\alpha\rangle_i\big) \\\nonumber
&=& \prod_i \big(\sum\limits_\alpha r_\alpha \exp ( i\alpha\phi) t_{\alpha,i}^\dag \big)|\text{vac}\rangle 
\stackrel{!}{=} \prod_i b_{0,i}^\dag  |\text{vac}\rangle.
\end{eqnarray} 
The introduction of $b^\dag_{0,i}$ as the new vacuum creation operator implies the need for a redefinition of the remaining two degrees of freedom. Such a transformation is performed via a two-parameter unitary rotation, whose rotation angles are chosen such that the new operators fluctuate around the new vacuum state and do not feature expectation values (cf. Fig. \ref{Fluctuations}),
\begin{eqnarray}\label{ratation}
b_{\alpha,i}^\dag = (R_\theta R_\chi)_{\alpha\beta} t_{\beta,i}^\dag
\end{eqnarray}
with the explicit form of the rotation matrices
\begin{eqnarray}\label{RotMat}
R_\theta &=&  \left(
\begin{array}{ccc}
  {\cos \theta/2} & {0} & {\sin \theta/2\mathrm e^{2\mathrm i \phi}}\\
  {0} & {1} & {0}\\
  {-\sin \theta/2 \mathrm e^{-2\mathrm i \phi}} & {0} & {\cos \theta/2}
\end{array}
\right), \\\nonumber
 R_\chi   &=&  \left(
\begin{array}{ccc}
  {1} & {0} & {0}\\
  {0} & {\cos \chi/2} & {-\sin \chi/2\mathrm e^{\mathrm i \phi}}\\
  {0} & {\sin \chi/2 \mathrm e^{-\mathrm i \phi}} & {\cos \chi/2}
\end{array}
\right) .
\end{eqnarray}
A finite $\theta (\chi)$ corresponds to a finite amplitude in $|2\rangle (|1\rangle)$. The precise relation is
\begin{eqnarray}\label{amplitudesII}
r_0 = \cos \theta/2, \,\, r_1 = \sin \theta/2 \sin \chi/2, \,\, r_2 =  \sin \theta/2 \cos \chi/2.\nonumber\\
\end{eqnarray}
The strategy is to first rotate to the new mean field state by inverting the unitary matrix in Eq. \eqref{ratation} and \emph{subsequently} implement the constraint. Analogous to the procedure in the physical vacuum, we may now eliminate the operator $b_0$ which is chosen to include the expectation value. We note that the local rotating frame transformation \eqref{RotFrame} can be applied also here, showing that the phase of $b_0$ is irrelevant for the Hamiltonian. Consequently we can implement the constraint by the formal replacement
\begin{eqnarray}\label{REP}
 b_{0,i}  \to X_i \equiv 1-  b_{1,i}^\dag  b_{1,i} -  b_{2,i}^\dag  b_{2,i},\quad 
b_{0,i}^\dag b_{0,i} \to X_i .
\end{eqnarray}
The second expression is simply a rearrangement of the holonomic constraint. The resulting bosonic Hamiltonian, which is then quantized by means of a functional integral, is rather complex, and its explicit form and analysis are discussed in \cite{Diehl09II}. However, it exhibits a simple structure,\begin{eqnarray}
H = E_{\text{GW}} +  H_{\text{SW}} + H_{\text{int}}.
\end{eqnarray}  
$E_{\text{GW}}$ is the Gutzwiller mean field energy and $H_{\text{SW}}$ describes the quadratic spin wave theory \footnote{A linear contribution, as naively expected in the expansion about the condensate, does not occur due to Goldstone's theorem, see Ref. \cite{Diehl09II}.}. The corrections to the mean field phase diagram, as well as nontrivial effects in the deep infrared physics which we analyze in \cite{Diehl09II} are not captured at this quadratic level. They are all encoded in the interaction part $H_{\text{int}}$.

At this point, let us compare our findings to the work of Huber \emph{et al.} \cite{Altman07}. We have verified explicitly that the quadratic part of the Hamiltonian coincides with the spin wave or Bogoliubov theory obtained in that work, though the authors use a different prescription for the resolution of the constraint for a single $b_0$ operator. The reason is that whenever the operator $b_0$ appears in multiplicative combination with another operator $b_1,b_2$, the replacement prescribed by the first expression of Eq. \eqref{REP} gives rise to at least cubic terms neglected in the spin wave approximation. In contrast, when the combination in the second expression of Eq. \eqref{REP} appears, one simply has to rearrange the constraint and no difference between the two approaches appears.

The many-body problem is completely specified only upon indicating conditions which determine the two rotation angles $\theta, \chi$. In \cite{Diehl09II} we show that they are fixed by gap equations which emerge as a consequence of Goldstone's theorem. Furthermore, the chemical potential is fixed via the equation of state when calculations at fixed filling are intended. These conditions are exact but implicit, and can be resolved approximately only. The simplest approximation reproduces the classical or mean field solution as anticipated above, but it is possible to go beyond this simple scheme with our setting \cite{Diehl09II}.

\subsection{Relation to a Spin-1 Model}
\label{sec:spin1}

From the availability of only three onsite states, it is clear that there exists a mapping of the 3-body constrained Bose-Hubbard model to a spin-1 model. Nevertheless, it seems advantageous to us to work with the above construction of mapping to a coupled boson theory. The reason is twofold. 

First, the practical analytical analysis of spin models beyond the standard mean field plus non-interacting spin wave approximation is technically very hard. Indeed, usually one resorts to introducing an artificial smallness parameter (inverse number of field components $1/N$, inverse total spin $1/S$), such that a Gaussian theory becomes exact in the limit $N,S\to \infty$, and organizes an expansion scheme around this noninteracting fixed point ($1/N,1/S$ expansions). This is then followed by a continuation of the results to the physical system of interest, where $N$ and $S$ are typically small. However, in our model $S=1$, and most of the effects which we discuss here and in \cite{Diehl09II} -- ranging from the non-perturbative formation of the dimer bound state in vacuum over corrections to the phase boundary to the true nature of the phase transition -- are not accessible to leading order in the abovementioned schemes, since they are all rooted in the intrinsic non-linearities of the theory. 

Second, typically the direct mapping of a bosonic theory with hardcore constraint  yields a rather complicated effective microscopic spin Hamiltonian, which further complicates an analysis in terms of spin degrees of freedom. For example, in our case the corresponding spin model would feature cubic and quartic bilocal spin interactions $\sim s_i^z s_i^+ s_j^- + {\rm h.c.}, \sim s_i^z s_i^+ s_j^- s_j^z + {\rm h.c.}$ with interaction constants of the same order as the quadratic terms \cite{Altman07}. These terms break the rotation symmetries typically present in generic Heisenberg models for magnets. Physically, the appearance of such terms has to be expected, since none of these rotation symmetries are present in the hardcore boson model. Clearly, mapping the constrained model to a theory of unconstrained coupled boson degrees of freedom, which find a natural interpretation in terms of single particle and bound state degrees of freedom, is closer to the physics of the boson theory with hardcore constraint. 

Reversely however, we emphasize that the class of spin models which can be readily encompassed within our formalism is made up of Heisenberg XX models in external fields, with possible extensions to XXZ models with small anisotropy. The power of conventional field theory techniques can thus be applied to such models.

\section{Fluctuations in the Vacuum Problem}
\label{VacProb}

In this section, we analyze the quantum field theory derived above in the limits $n=0,2$. This provides a useful starting point for the treatment of the many-body problem addressed in \cite{Diehl09II}. First we focus on the two-body problem, for which we present the exact solution within our framework. The Schr\"odinger equation for two-particle scattering is correctly reproduced. Clearly, for two particles, the physical three-body constraint cannot play a role. However, a mathematically wrong or only approximate implementation of the constraint will produce a wrong scattering equation as we see explicitly. Second, we use the language of Feynman diagrams to explicitly calculate the two-dimer interaction strength up to fourth order in the perturbative regime $J/|U|\ll 1$. This provides another benchmark for our formalism.  Third, we consider hole scattering in the limit $n=2$ and discuss the formation of a di-hole bound state, which exhibits properties different from the dimer bound state at $n=0$. 

Below we introduce the formalism used to do concrete calculations, the Dyson-Schwinger equations, demonstrating how to use the powerful methods of quantum field theory in our problem. These equations provide an exact hierarchy of relations between correlation functions.  We find that in the vacuum limit, where the system of equations describes few-particle scattering, the equations for the dimer self energy and the splitting vertex are one-loop. Furthermore, the atom self energy is not renormalized. These three generalized couplings form a closed system of equations, decoupling from higher interaction vertices. We find the exact solution for these equations. These ingredients lead to the exact solution of the two-body problem, which manifests itself in the emergence of the non-perturbative Schr\"odinger  equation for the bound state. 

In order to compute higher interaction vertices, we need to take higher loop diagrams into account, and the system of equations for these vertices is not closed. However, we can establish a perturbative expansion of the equations in the limit $J/|U|\to 0$. In \cite{Diehl09II}, we establish the relation of the resulting effective theory to a spin-1/2 model in this limit. 


\subsection{Dual Feshbach Model}
\label{sec:DualFesh}

Before embarking the calculations, let us briefly discuss the microscopic model emerging from our quantization prescription. We start with the microscopic action, obtained from the Hamiltonian Eq. \eqref{HC} using Eq. \eqref{ActHam}. The complex fields $t^\dag_{1,2},t_{1,2}$ are now fluctuating classical variables and we may permute them at will,
\begin{widetext}
\begin{eqnarray}\label{FinalAction}
S[t_1,t_2] &=& \int d \tau\Big( \sum\limits_i\Big[  t_{2,i}^\dag(\tau)(\partial_\tau - 2\mu + U) t_{2,i}(\tau) + t_{1,i}^\dag(\tau)(\partial_\tau - \mu ) t_{1,i}(\tau)\Big] \\\nonumber
&&  \qquad - J \sum_{\langle i,j\rangle} \Big[ t_{1,i}^\dag (\tau)  t_{1,j}(\tau)X_i(\tau)X_j(\tau)   + 2 t_{2,i}^\dag (\tau)t_{2,j}(\tau)t_{1,j}^\dag(\tau) t_{1,i}(\tau)\\\nonumber
&&\qquad   + \sqrt{2} \big[ t_{2,i}^\dag(\tau)  t_{1,i}(\tau) t_{1,j}(\tau) + t_{2,i} (\tau) t_{1,i}^\dag(\tau) t_{1,j}^\dag(\tau)\big] X_j(\tau)\Big]   \Big).
 \end{eqnarray}
 \end{widetext}
This action has the form of a Feshbach resonance model (see \cite{Holland01} for the fermion case, and \cite{Radzihovsky04,Sachdev04} for the bosonic case) on the lattice, with non-local interaction parameters: The dimer degree of freedom couples to the chemical potential with double strength, taking care of the double atom number in the dimer. The role of the "detuning" of the dimer state from the atoms is played by the onsite interaction. We note an interesting duality to the standard Feshbach model in the continuum for atom and dimer degrees of freedom obtained from a Hubbard-Stratonovich decoupling of an attractive (local) two-body interaction $u$: Such a procedure generates a detuning $\nu \sim 1/u$, while in our case the detuning $\nu \sim U$. Physically, this means that the bound state formation in our lattice scenario is a weak coupling phenomenon, while being a strong coupling (resonant) effect in the continuum. Further note that the usual lattice construction using the single band approximation is delicate for systems close to Feshbach resonances \cite{Duan05}, and a realization of lattice Feshbach models in the resonant case is therefore not a straightforward task, while it is realized in a natural way here.  

The physically most important coupling is the Feshbach or splitting vertex. It describes the formation of a dimer out of two atoms and its reverse. In contrast to the conventional continuum Feshbach model, the splitting vertex in our model is bi-local, which in momentum space induces a form factor but does not lead to complications and still allows for an exact solution of the two-body problem. The term in the third line describes a non-local interaction between atoms and dimers. 

In the loop calculations we prefer to work in frequency and momentum space. With the definitions 
\begin{eqnarray}
t_{\alpha, i}(\tau) &=& \int_q e^{\mathrm i q x_i} t_{\alpha, q}, \quad t^\dag_{\alpha, i}(\tau) = \int_q e^{-\mathrm i q x_i} t^\dag_{\alpha, q} ,\\\nonumber
 x_i &=& (\tau, \textbf{x}_i),\quad q = (\omega,\textbf{q}), \quad \int_q = \int\frac{d\omega}{2\pi} \sum_\textbf{q}, \\\nonumber
 \epsilon_\bq\hspace{-0.15cm} &=&\hspace{-0.15cm} J\sum\limits_{\lambda=1}^d\cos (\bq { \textbf{e}}_\lambda ),\,\ \Delta X_{q,k} = +( t^\dag_{1,q}t_{1,k} + t^\dag_{2,q}t_{2,k}) 
\end{eqnarray}
the Fourier transformed action reads
\begin{widetext}
\begin{eqnarray}\label{FourierAct}
S[t_1,t_2]  &=& \int\limits_q  \Big[ t_{1, q}^\dag (\mathrm i \omega  - \mu - 2\epsilon_\bq ) \,\,t_{1, q} + t_{2, q}^\dag (\mathrm i \omega  - 2\mu +U) \,\,t_{2, q}  \Big]\\\nonumber
&& -\sqrt{2}  \int\limits_{q_1,q_2,q_3} \delta (q_1 -q_2 - q_3) (\epsilon_{\bq_2}+\epsilon_{\bq_3}  )(t_{2, q_1}^\dag t_{1, q_2} t_{1, q_3} + \text{h.c.} ) \\\nonumber
&& - 2 \int\limits_{q_1, ... ,q_4} \delta (q_1 - q_2 + q_3 - q_4) [\epsilon_{\bq_1 - \bq_4} + \epsilon_{\bq_2 - \bq_3} ] t_{1, q_1}^\dag t_{1, q_2} t_{2, q_3}^\dag  t_{2, q_4} \\\nonumber
&& + \sqrt{2} \int\limits_{q_1, ... ,q_5} \delta (q_1 - q_2 - q_3 + q_4 - q_5 )[ \epsilon_{\bq_1 - \bq_2} + \epsilon_{\bq_1 - \bq_3}]( t_{2, q_1}^\dag  t_{1, q_2} t_{1, q_3} \Delta X_{q_4, q_5}+\text{ h.c.})\\\nonumber
&& +2 \int\limits_{q_1, ... ,q_4} \delta (q_1 - q_2 + q_3 - q_4 )[ \epsilon_{\bq_1 } + \epsilon_{\bq_2}] t_{1, q_1}^\dag  t_{1, q_2} \Delta X_{q_3, q_4}\\\nonumber
&& - \int\limits_{q_1, ... ,q_6} \delta (q_1 - q_2 + q_3 - q_4 + q_5 - q_6) [ \epsilon_{\bq_2 + \bq_3 - \bq_4} + \epsilon_{\bq_2 + \bq_5 - \bq_6} ]t_{1, q_1}^\dag t_{1, q_2} \Delta X_{q_3, q_4}\Delta X_{q_5, q_6} .
\end{eqnarray}
\end{widetext}

\subsection{Dyson-Schwinger Equations}
\label{sec:DSE}

Dyson-Schwinger equations (DSEs) \cite{Dyson49Schwinger51} are a direct consequence of the shift invariance of the functional integral:
\begin{eqnarray}\label{ShiftInvariance}
0  &=& \frac{1}{Z[j]}\int \mathcal{D}(\delta\hat\chi)\, \frac{\delta}{\delta \hat\chi}\exp - S[\hat\chi] + j^T\delta \hat\chi \\\nonumber
 &=& \frac{1}{Z[j]}\int \mathcal{D}(\delta\hat\chi)\, \Big(- \frac{\delta S}{\delta \hat\chi} + j\Big)^T \exp - S[\hat\chi]
 + j^T\delta \hat\chi.
\end{eqnarray}
Switching to the effective action, i.e. requiring $j =  \delta \Gamma/\delta \chi$, the above equation turns into
\begin{eqnarray}\label{SDEGen}
\frac{\delta \Gamma}{\delta \chi} = \Big\langle \frac{\delta S}{\delta \hat\chi} \Big\rangle
\Big|_{j =  \delta \Gamma/\delta \chi}.
\end{eqnarray}
This is the DSE for the one-point function. To reveal the structure of the DSEs for higher $N$-point functions, we consider a general classical action with $M$ vertices. We write the classical action in a vertex expansion, 
\begin{eqnarray}
S[\hat\chi] &=& S[\chi] + \sum\limits_{N = 1}^{M}\frac{1}{N!}  S^{(N)}_{\alpha_1 ... \alpha_N}\delta\hat\chi_{\alpha_1}
\cdot ... \cdot \delta\hat\chi_{\alpha_N}.
\end{eqnarray}
Here $\alpha_i$ is a multi-index collecting field type as well as space and timelike (or momentum and frequency) indices. $S$ and $S^{(N)}$ still depend on the classical field $\chi$. Plugging the vertex expansion into Eq. (\ref{SDEGen}) relates the field derivative of the effective action to 1PI Green functions up to order $M$. We can turn the DSE into a manifestly closed equation, i.e. an equation which is expressed solely in terms of the effective action and its functional derivatives. It reads 
\begin{eqnarray}\label{SDEGenII}
\frac{\delta \Gamma}{\delta \chi_\beta}\hspace{-0.2cm} &=& \hspace{-0.1cm}S^{(1)}_\beta 
+ \frac{1}{2!}S^{(3)}_{\alpha_1\alpha_2 \beta}G_{\alpha_1\alpha_2}\\\nonumber
&&\hspace{-0.3cm}+\hspace{-0.1cm} \sum\limits_{N = 4}^{M}\frac{1}{(N-1)!}
S^{(N)}_{\alpha_1 ... \alpha_{N-1}\beta}\Big[\prod\limits_{i=3}^{N-1}G_{\alpha_i\kappa_i}
\frac{\delta }{\delta \chi_{\kappa_i}}\Big]
G_{\alpha_1\alpha_2}.
\end{eqnarray}
For $N=4$ the derivative operator in the squared brackets is just the unit matrix. 
The full propagator is denoted by $G$ and we have the relation $G_{\alpha\beta} =( \Gamma^{(2)})^{-1}_{\alpha\beta}$. The full propagator as well as the classical vertices are functions of the classical field, $S^{(N)} = S^{(N)}[\chi], G = G[\chi]$, such that the DSE for the $N$-point correlation function can be obtained by taking $N-1$ functional derivatives on Eq. (\ref{SDEGenII}). We observe that the one-point function $\delta \Gamma / \delta \chi_\beta$ depends on correlation functions up to order $M-1$. Thus, the DSE for the $N$-point function features vertices up to order $M + N -2$. Furthermore the self-consistency equations for the correlation functions for a theory with classical vertices up to order $M$ features $M-2$-loop diagrams, since the order $M$ vertex has $M-1$ internal lines.

\subsection{Exact Solution of the Two-Body Problem}
\label{VacProbSol}

\begin{figure}
\begin{center}
\includegraphics[width=0.9\columnwidth]{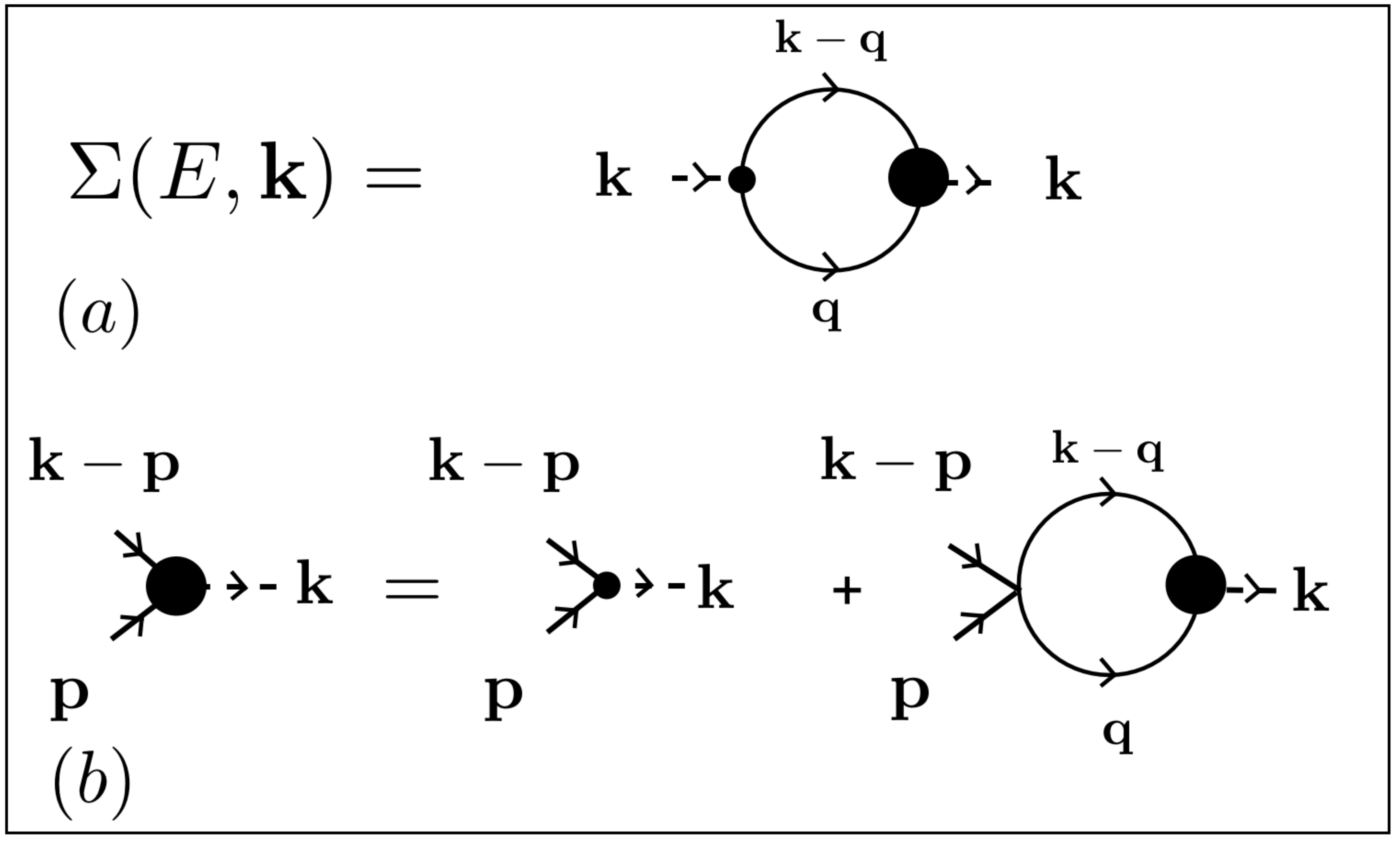}
\end{center}
\caption{\label{fig1} Scattering equations in vacuum. The solid lines represent atom propagators. Full vertices are signalled with heavy blobs. (a) Equation for the dimer self-energy. The full Feshbach vertex is needed for its solution. (b) Renormalization of the Feshbach vertex. External momentum configurations are chosen as needed in (a). 
}
\end{figure}

\begin{figure}
\begin{center}
\includegraphics[width=0.9\columnwidth]{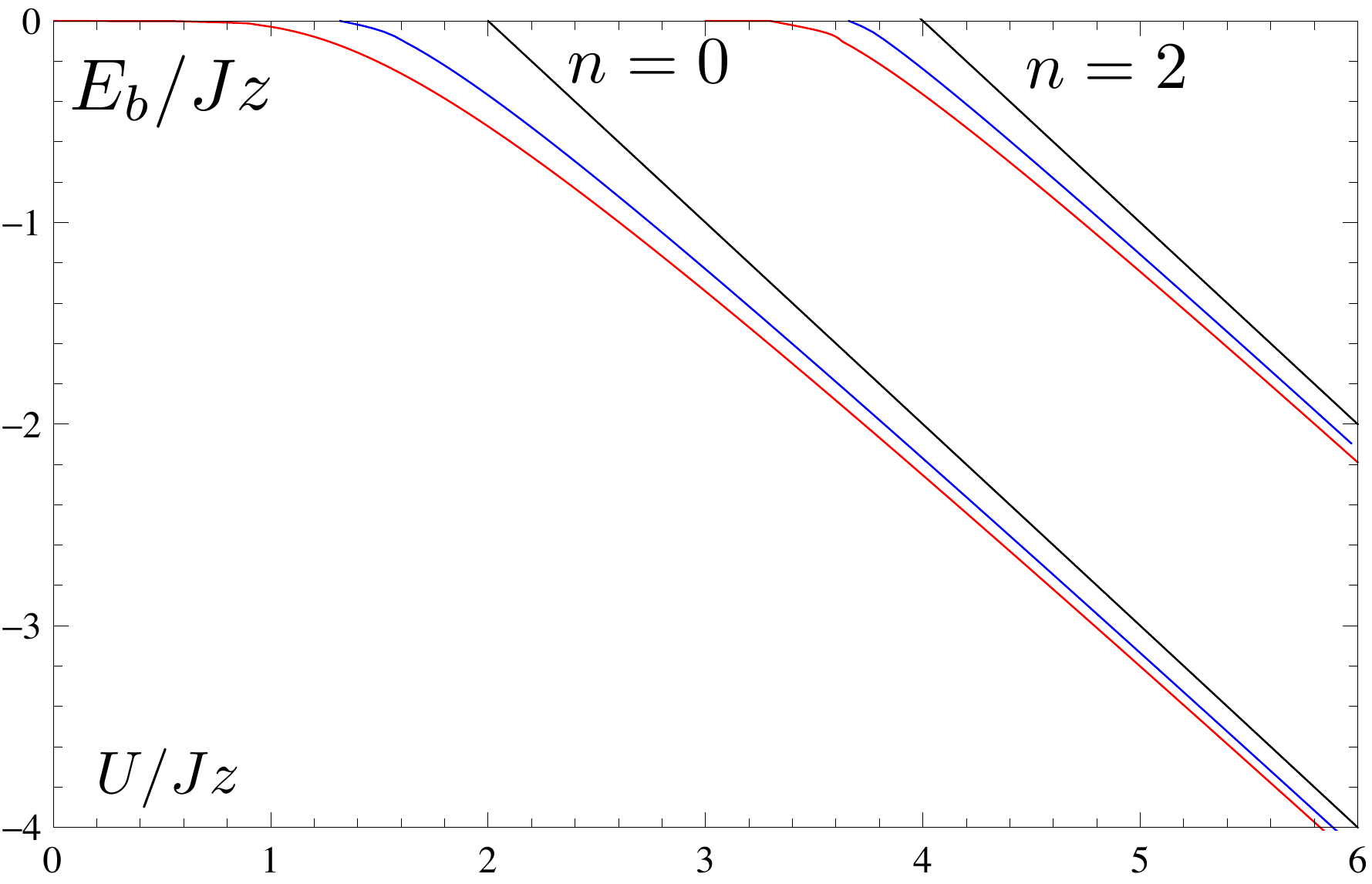}
\end{center}
\caption{\label{BindingEnergy} Dimensionless binding energy of dimers ($n=0$) and di-holes ($n=2$) as a function of the dimensionless interaction strength. The upper lines (black online) denote the mean field results, the lower/middle curves (red/blue online) are the exact binding energies in dimensions 2 and 3, respectively.  }
\end{figure}

The scattering problem is described by two coupled integral equations for the exact dimer self-energy and for the exact  Feshbach vertex, cf. Fig. \ref{fig1} and App. \ref{app:DSE} for the derivation. Inserting the frequency and momentum configurations appropriate for two-body scattering as depicted in Fig. \ref{fig1} and integrating out the frequencies, we obtain
\begin{eqnarray}\label{ScattVac}
G_d^{-1}(E,\textbf{k}) \hspace{-0.1cm}&=&\hspace{-0.1cm}  G_d^{(0)\,-1}(E,\textbf{k}) +\Sigma(E,\textbf{k}), \nonumber\\
\Sigma(E,\textbf{k}) &=&-  \frac{1}{\sqrt{2}} \int\frac{d^dq}{(2\pi)^d} \frac{\Gamma_\textbf{k}(\bq)\Gamma^{(0)}_\textbf{k}(\bq) }{ E + \Gamma^{(0)}_\textbf{k}(\bq)  }, \nonumber\\
\Gamma_\textbf{k}(\textbf{p})  \hspace{-0.1cm}&=&\hspace{-0.1cm} \Gamma^{(0)}_\textbf{k}(\textbf{p})  + \hspace{-0.15cm}\int\hspace{-0.1cm}\frac{d^dq}{(2\pi)^d} \frac{\Gamma_\textbf{k}(\bq) (\Gamma^{(0)}_\textbf{k}(\textbf{p})  +\Gamma^{(0)}_\textbf{k}(\bq) )}{ E + \Gamma^{(0)}_\textbf{k}(\bq)  } ,\nonumber\\
\end{eqnarray}
with the definitions 
\begin{eqnarray}
\Gamma^{(0)}_\textbf{k}(\bq) = -2\sqrt{2} (\epsilon_\bq + \epsilon_{\bq - \textbf{k}}), \quad E = \sqrt{2}( \mathrm i \omega - 2\mu ),
\end{eqnarray}
where $\omega$ is the Euclidean external frequency. The difference between the full ($G_d(E,\textbf{k})$) and the bare ($G_d^{(0)}(E,\textbf{k})$) Green's function is the dimer self-energy, $\Sigma(E,\textbf{k}) = G_d^{-1}(E,\textbf{k}) - G_d^{(0)\,\,-1}(E,\textbf{k})$. Note carefully the appearance of both external ($\textbf{p}$) and loop ($\bq$) momenta in the equation for the full Feshbach vertex. The second equation can be solved independently of the first one. This can be done by choosing the following ansatz for the full vertex,
\begin{eqnarray}
\Gamma_\textbf{k}(\textbf{q }) = \Gamma^{(0)}_\textbf{k}(\textbf{q}) \gamma^{(0)}(E,\textbf{k}) + \gamma^{(1)}(E,\textbf{k}),
\end{eqnarray}
where the unknown $\gamma^{(0)}$ is dimensionless while $\gamma^{(1)}$ carries dimension of energy. The two unknown functions depend only on the external center-of-mass momentum $\textbf{k}$ and the energy variable $E$; the dependence on the relative momentum $\bq$ only appears in the coefficient of $\gamma^{(0)}$. Comparing coefficients this ansatz yields the following system of coupled equations for the unknowns
\begin{eqnarray}
\gamma^{(0)} &=& 1 +  \gamma^{(0)} (1 - E I )  + \gamma^{(1)} I, \\\nonumber
\gamma^{(1)} &=&  [\gamma^{(1)} - \gamma^{(0)} E ](1 - E I ),
\end{eqnarray}
where we use the abbreviation 
\begin{eqnarray}
I (E,\textbf{k}) =  \int\frac{d^dq}{(2\pi)^d} \frac{1}{ E + \Gamma^{(0)}_\textbf{k}(\bq)  } 
\end{eqnarray}
and the simplifications
\begin{eqnarray}
&& \int\frac{d^dq}{(2\pi)^d} \frac{\Gamma^{(0)}_\textbf{k}(\bq) }{ E + \Gamma^{(0)}_\textbf{k}(\bq)  }  = 1 - E I ,\\\nonumber
&&  \int\frac{d^dq}{(2\pi)^d} \frac{\Gamma^{(0)}_\textbf{k}(\bq) \Gamma^{(0)}_\textbf{k}(\bq)}{ E+ \Gamma^{(0)}_\textbf{k}(\bq)  }  = E(  E I - 1).
\end{eqnarray} 
The solution of the above equations is 
\begin{eqnarray}\label{FullYukawa}
\gamma^{(0)}  = 1, \qquad \gamma^{(1)}  = E - I^{-1} ,\\\nonumber
\Gamma_\textbf{k}(\bq)  = (E + \Gamma^{(0)}_\textbf{k}(\bq) ) - I^{-1}.
\end{eqnarray}
The exact self-energy is given by 
\begin{eqnarray}
\Sigma(E,\textbf{k}) &=& \frac{1}{\sqrt{2}} ( I^{-1} - E) = - (\mathrm i  \omega -2 \mu  ) +\\\nonumber
&& \Big[ \int\frac{d^dq}{(2\pi)^d} \frac{1}{- 2 (\epsilon_\bq  + \epsilon_{\bq - \textbf{k}}) +\mathrm i  \omega - 2\mu  } \Big]^{-1},
\end{eqnarray}
such that the equation for the full inverse dimer Green's function becomes, with $G_d^{(0)\,-1}(E,\textbf{k}) = \mathrm i \omega -2\mu + U$, 
\begin{eqnarray}\label{FullGreen}
G_d^{-1} (E,\textbf{k}) = U + \Big[ \int\frac{d^dq}{(2\pi)^d} \frac{1}{- 2 (\epsilon_\bq  + \epsilon_{\bq - \textbf{k}})  +\mathrm i  \omega - 2\mu} \Big]^{-1}.\nonumber\\
\end{eqnarray}
The presence of a bound state is signalled by a pole in the dimer Green's function at zero center-of-mass momentum and zero external frequency, $G_d^{-1}(\omega =0; \mu , \textbf{k}=0) =0$. The chemical potential $\mu$ in the physical vacuum can be interpreted as the binding energy \cite{Diehl06} after an appropriate decomposition which ensures that the atom at rest has no kinetic energy, $\mu = \mu_b- Jz$. This definition separates true kinetic from true potential (binding) energy, and finite momentum excitations have positive energy, $\delta\epsilon_\textbf{q} = J \sum_\lambda (1 - \cos \bq \textbf{e}_\lambda )$. Then, we have for the binding energy $E_b = 2\mu_b$, and we note that the atoms are gapped out with half the binding energy, since their Green's function involves $-\mu_b$. Thus, the molecular degrees of freedom are the lowest excitations since in contrast to the atoms they are massless (pole condition). For vanishing binding energy we have $\mu_b=0$, such that the situation is reversed: The atoms are the gapless excitations, while the molecules are gapped. Introducing dimensionless and dimensionally invariant variables $\tilde U = U/(Jz), \tilde E_b = E_b /(Jz)$ the pole condition leads to 
\begin{eqnarray}\label{LSE3}
\frac{1}{| \tilde U| } = \int\frac{d^dq}{(2\pi)^d} \frac{1}{ - \tilde E_b + 2/d \sum_\lambda (1 - \cos \bq \textbf{e}_\lambda ) } .
\end{eqnarray}
This is precisely the Schr\"odinger equation for the dimer bound state. We discuss the formation of the bound state, since it will be interesting to confront these well-known results to the formation of di-hole bound states at $n=2$ which shows different properties. The bound state forms at the critical $\tilde U_d$ in $d$ dimensions where $E_b=0$. In three dimensions, the integral evaluates to a finite value, 
 while in two dimensions a logarithmic infrared divergence pushes $\tilde U_2$ to zero:
\begin{eqnarray}
\tilde U_3 \approx  - \frac{4}{3}, \quad \tilde U_2 = 0.
\end{eqnarray}
This has to be compared to the mean field result at zero density, $\tilde U_{\mathrm{mf}} =-2$, obtained from the classical inverse dimer Green's function $G^{(0)\,-1}_d(\omega =0;\mu,\textbf{k} =0)$. We observe a substantial downshift in the critical interaction strength. Close to the onset of the bound state the binding energy starts quadratically ($d=3$) resp. exponentially ($d=2)$, due to the square root nonanalyticity resp. logarithmic divergence of the fluctuation integral: 
\begin{eqnarray}\label{QuadraticBinding}
d= 3 &:& \tilde E_b  \approx - \Big(\frac{|\tilde U| - |\tilde U_3| }{\sigma \tilde U_3}\Big)^2,\\\nonumber 
d=2 &:& \tilde E_b  \approx -\tfrac{\Lambda^2}{2} \exp (- \tfrac{2\pi}{| \tilde U|}) ,\\\nonumber
&& \tilde E_{b,\mathrm{mf}} = \tilde U_{\mathrm{mf}} - | \tilde U |   = 2 - | \tilde U |   ,
\end{eqnarray}
with numbers $\sigma \approx 0.42, \Lambda \approx 5.50$ determined numerically. We have added the linear mean field result. This dimensionally invariant behavior is approached for large negative couplings, cf. Fig. \ref{BindingEnergy}.

The fact that the Schr\"odinger  equation is reproduced by our nonperturbative calculation is an important benchmark for our theory. Note that any deviating implementation of the constraint, such as an expansion of the square root Eq. \eqref{NaiveConstraint} to leading order, would generate incorrect prefactors in the scattering equations, and the Schr\"odinger  equation could not be reproduced. 
\vspace{0.5cm}

\subsection{Perturbative Limit: Effective Two-Body Hard Core Dimer Gas}
\label{sec:EFT}

While the perturbative limit for the two-body problem is straightforwardly obtained to any order from the exact solution above, for higher order vertices, nonperturbative calculations are hard -- the computation of the full dimer-dimer scattering vertex would require the complete solution of the four-body problem. However, in the limit $J/U \ll 1$ a systematic perturbative expansion of our set of Dyson-Schwinger Equations is available. For example, at order $J^2$ only diagrams with at most two vertices have to be taken into account. We find that a loop expansion to one-loop order is insufficient even at order $J^2$. We may understand that qualitatively from the fact that on short ranges, we are dealing with a full quantum mechanical problem with strong fluctuations -- the occupation of a site is either 0 or 1. Thus, a loop expansion (in orders of $\hbar$) cannot be expected to be reliable.

In this section, we calculate the effective Hamiltonian for dimers. There dimers will obviously have a two-body hardcore constraint, as they are made up of two atoms each. The relation to a spin 1/2 model is made in \cite{Diehl09II}. There, we also show that a symmetry enhancement from the conventional $U(1) \sim SO(2)$ for bosons to $SO(3)$ is taking place in the strongly correlated limit, and discuss its physical implications. 

We will first calculate the effective Hamiltonian up to second order perturbation theory. We then perform a partial calculation of the fourth order. These results are crucially needed when discussing the many-body phases in the strongly coupled regime in \cite{Diehl09II}, as well as for the discussion of the nature of the phase transition. 

\subsubsection{Second Order}

We are interested in extracting the effective action for dimers in the perturbative limit. If restricting to second order in $J$, the effective theory can only contain nearest neighbour terms. The most general form compatible with the constraint and symmetry principles is given by 
\begin{widetext}
\begin{eqnarray}\label{SupplementAction}
S[ t_2] &=& \int d \tau\Big( \sum_{i} Z t_{2,i}^\dag(\tau)( \partial_\tau   - \mu_d) t_{2,i}(\tau)  - t \sum_{\langle i,j\rangle} t_{2,i}^\dag(\tau)  X_i(\tau)X_j(\tau) t_{2,j}(\tau)+\frac{v}{2} \sum_{\langle i,j\rangle} Ê\hat n_{2,i}(\tau) \hat n_{2,j} (\tau)   \Big).
\end{eqnarray}
\end{widetext}
The $X$ terms are introduced in order to satisfy the constraint principle, and there cannot be an onsite dimer interaction for the same reason. $Z$ is a wave function renormalization factor. It accounts for the energy dependence of the perturbative expansion in $x=J/|U|$. $\mu_d = 2\mu - U + \Delta \mu_d \leq 0$ is the effective dimer chemical potential. $t, v$ are the constrained hopping and interaction constants to be determined. 

\emph{Kinetic Terms} -- The desired information on the local and the constrained hopping term can be extracted from the solution of the two-body problem, Eq. \eqref{FullGreen}. For this purpose we take $\mu \to -\infty$,
\begin{eqnarray}\label{ApproxGreen}
G_d^{-1} (\omega;\mu ,\textbf{k}) = \mathrm i \omega  - 2\mu  + U  -  \frac{4 J^2}{\mathrm i \omega  - 2\mu} \sum_{\lambda} ( 1 + \cos \textbf{k} \textbf{e}_\lambda ).\nonumber\\
\end{eqnarray}
In this form it is apparent that the full inverse propagator contains the microscopic one exactly, but is now supplemented by a qualitatively new hopping term. To be consistent at second order, in the denominator we have to insert $\mu = -|U|/2$. After Fourier transformation of the quadratic part of the action with inverse Green function \eqref{ApproxGreen}, we obtain $\Delta \mu_d  =  2J^2z/|U|, t = 2J^2z/|U|$.
The wave function renormalization factor $Z$ is extracted from Eq. \eqref{ApproxGreen} from expanding in i$\omega$; there is no second order contribution at second order perturbation theory, i.e. $Z=1$.

\begin{figure}
\begin{center}
\includegraphics[width=0.9\columnwidth]{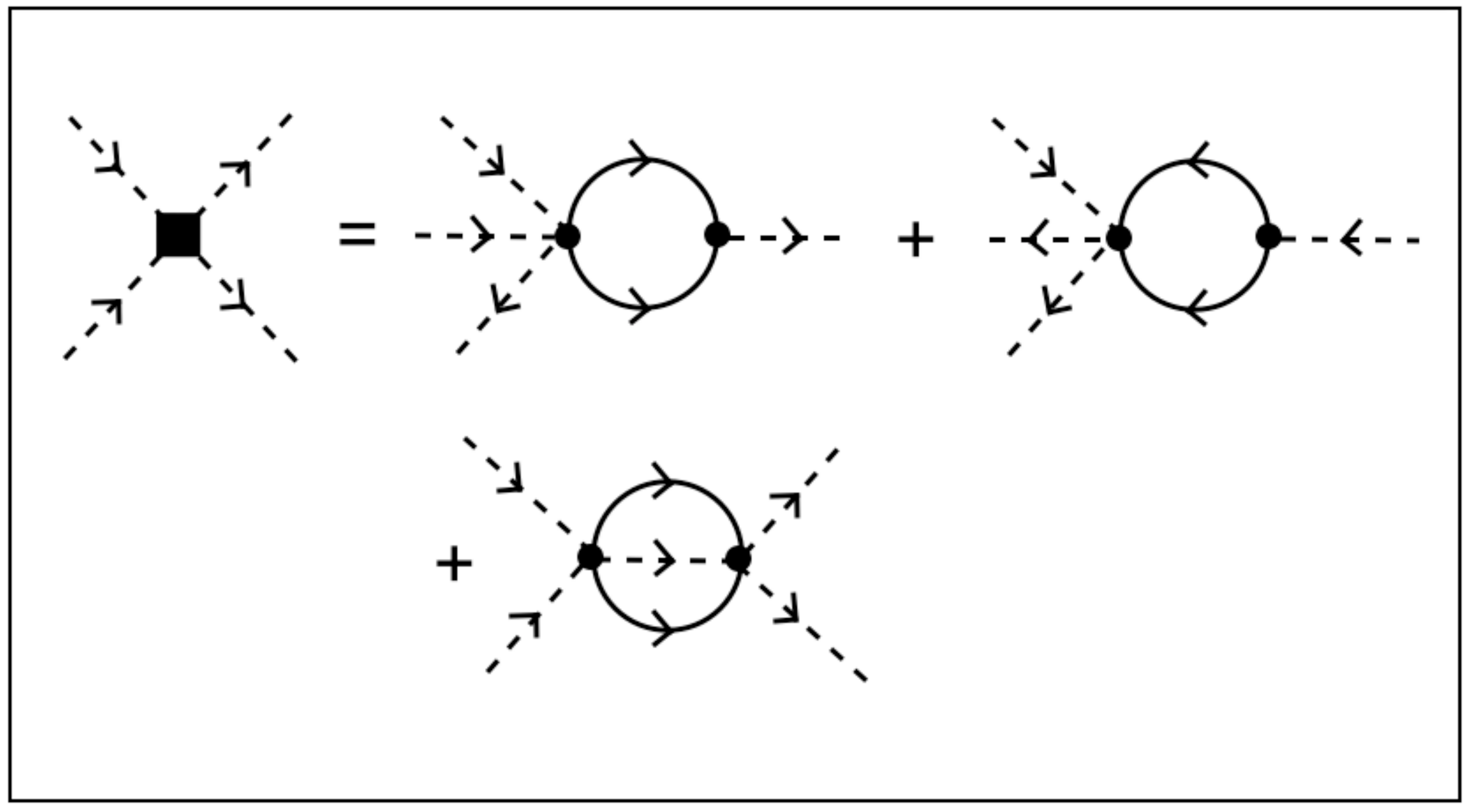}
\end{center}
\caption{\label{DimerDimer} The dimer-dimer interaction vertex to second order in the hopping $J$.}
\end{figure}

\emph{Dimer-Dimer Interaction} -- At short ranges, one expects a dimer density-density repulsion due to the reduced decay and recombination possibilities of one dimer if there is another one sitting close by. The Dyson-Schwinger Equation governing the dimer-dimer scattering in the perturbative regime is displayed in Fig. \ref{DimerDimer}. The one-particle irreducible graphs give rise to a nearest-neighbour (second order in $J$) density-density repulsion, due to the presence of the constraint, manifesting itself via the five-point splitting vertex. As anticipated above, it is interesting to note that the equation is two-loop even in the leading order perturbation theory. For the derivation of the symmetry factors and the explicit calculation we refer to App. \ref{app:DSE}. Here we indicate the second order result of the momentum space calculation and discuss it: 
\begin{eqnarray}\label{DimerDimerInteractionEx2}
\frac{v_{\{ \textbf{k}_i\}}}{2}\hspace{-0.1cm}=\hspace{-0.05cm} \frac{2J^2}{|U|} \big( \epsilon_{ \textbf{k}_1 - \textbf{k}_2 } \hspace{-0.05cm}+\hspace{-0.05cm} \epsilon_{ \textbf{k}_3 - \textbf{k}_4 } \hspace{-0.05cm}+\hspace{-0.05cm} \epsilon_{ \textbf{k}_1} \hspace{-0.05cm}+\hspace{-0.05cm} \epsilon_{\textbf{k}_2} +  \epsilon_{ \textbf{k}_3} + \epsilon_{\textbf{k}_4}\big).
\end{eqnarray}
The two qualitatively different momentum dependences correspond to different interaction processes in position space. After Fourier transform we find the contribution to the effective action
\begin{eqnarray}
 \frac{2 J^2}{|U|}\hspace{-0.1cm} \int\hspace{-0.15cm} d \tau  \hspace{-0.1cm}\sum_{\langle i,j\rangle}  \Big[Ê \hat n_{2,i}(\tau) \hat n_{2,j} (\tau) +  t_{2,i}^\dag(\tau) t_{2,j}(\tau) \big(\hat n_{2,i} + \hat n_{2,j} \big) \Big].\hspace{-0.3cm}\nonumber\\
\end{eqnarray}
The terms may be interpreted as follows. The first term is a true dimer-dimer interaction describing the exchange of dimers on adjacent sites and is repulsive. The second one is the explicit manifestation of the constraint being inherited by the effective theory of dimers -- they precisely contribute the terms linear in $\hat n_2$ which are contained in $X_iX_j$ in Eq. \eqref{SupplementAction}.

In summary, the effective couplings to second order read
\begin{eqnarray}
 Z =1, \quad \Delta \mu_d = t =  \frac{v}{2} =   \frac{2 J^2}{|U|} .
\end{eqnarray}
Both effective hopping and interaction are not present in the mean field approximation. Conceptually and practically, they are however of high importance. The first one makes the dimers true physical, i.e. spatially propagating degrees of freedom, while the second one, with the positive sign, is very important for the many-body and long-wavelength calculations carried out in \cite{Diehl09II}, as it stabilizes the thermodynamic potential for the dimer superfluid. It also gives rise to the stiffness of the superfluid.  Since the second order contributions are due to fluctuations on a single link of nearest neighbours, the results are dimension independent.

\subsubsection{Fourth Order}

At fourth order, we obtain not only nearest neighbour (nn) contributions to the constrained hopping and interaction, but also next-to-nearest neighbour (nnn) contributions. However, due to the reduced number of pathways connecting  these more distant sites, the coefficients are substantially smaller then for nn terms. In the following, we thus concentrate on the latter. In particular, for later purposes we will be interested in the deviation of the ratio of interaction vs. kinetic energy $\lambda = v/(2t)$ from the second order result $\lambda =1$. 

A brute force diagrammatic fourth order calculation is rather complex. Here we present a way to perform the fourth order calculation of the nn coeffcients for the interaction from a combination of geometric and diagrammatic arguments: First we argue that based on the geometry of the contributing processes, the repulsive part of the fourth order contribution must equal the fourth order hopping contribution. The latter, in turn, can be calculated straightforwardly from the exact solution of the two-body problem. This argument can be applied for processes of arbitrary intersite distance, and we will present it in its general form. Then we refocus on nearest neighbours and identify processes contributing to the interaction which have no analog in the hopping process. We calculate the corresponding reduced set of diagrams explicitly. They yield an attractive contribution to the interaction strength such that $\lambda <1$. 

We begin with the geometric argument and consider the hopping first. If we are interested in the energetic contribution of hopping processes at a fixed order in perturbation theory in $J/|U|$ and at a fixed distance, we can find this contribution in principle by drawing all possible pathways with a fixed number of hops which connect initial and final site. We call these paths connecting initial and final state via hopping the hopping paths. Next we consider the interaction coefficient. The physical origin of this interaction emerges from the constraint: if there is a dimer at site $j$, then another particle cannot hop on this site. This can be calculated as the energy difference that emerges when comparing the number of paths through which a dimer at site $i$ can decay and come back to this site without a dimer sitting at $j$, and the corresponding number when there is a dimer on $j$. The energy difference obviously is always positive, because the constraint always excludes a stet of paths, leading to a repulsive interaction. Thus, the energy contribution can be obtained by just counting the number of paths where at least one of the travelling particles hits the site $j$. We call these paths interaction paths. 

Now we observe that any hopping path can be transformed into an interaction path by reversing the direction of the arrows of one of the travelling particles \emph{on the shortest path} which connects initial and final site (there may be several of these shortest paths) for any overall length of the paths, i.e. at any order of perturbation theory. Therefore, the number of interaction paths is larger or equal compared to the number of hopping paths. The reverse is also true. Thus the number of interaction and hopping paths is equal, and therefore the so-obtained contribution to the interaction must equal the fourth order hopping contribution. 

\begin{figure}
\begin{center}
\includegraphics[width=0.9\columnwidth]{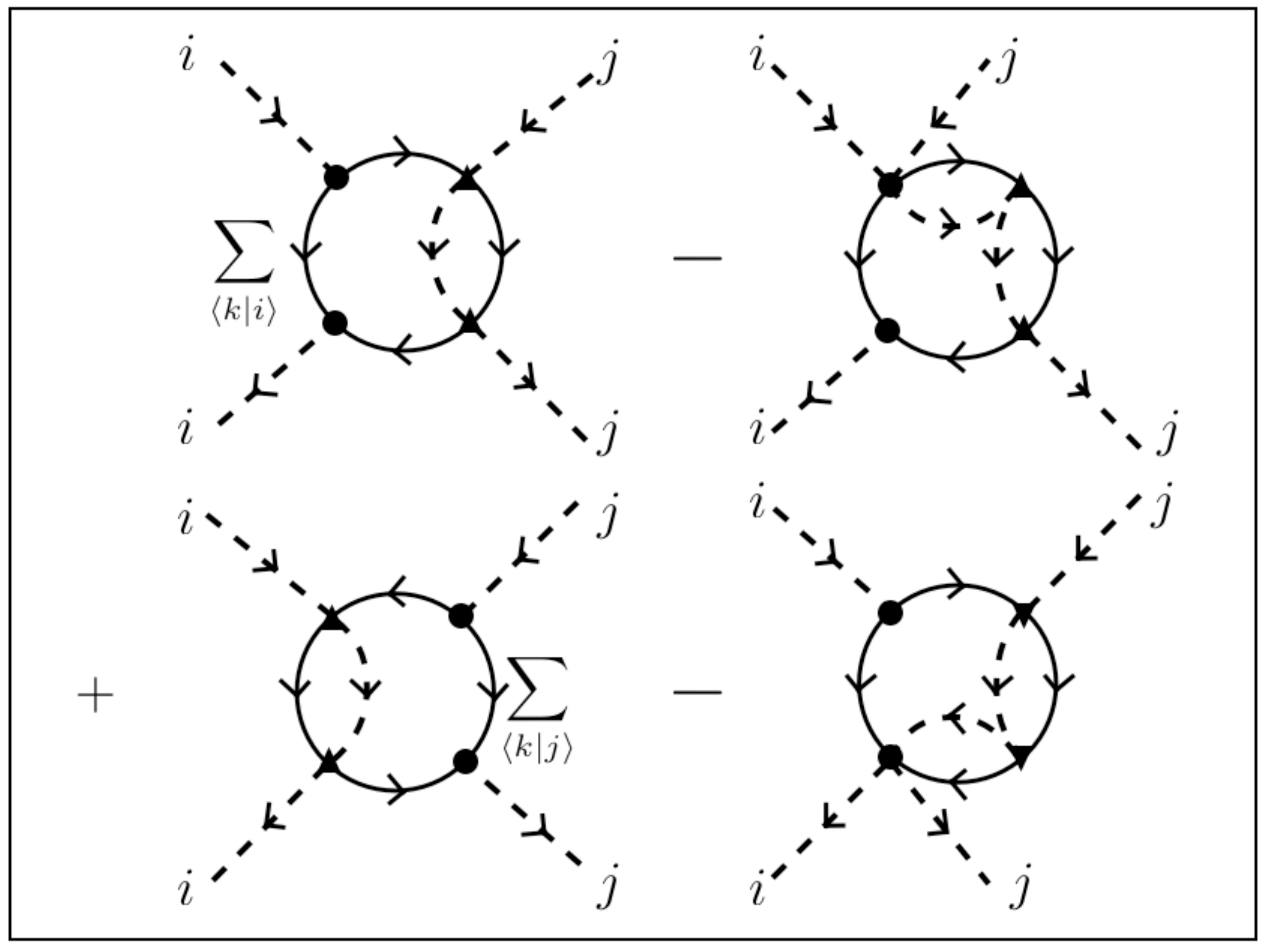}
\end{center}
\caption{\label{TwoLoopAttractive} Attractive contribution to the nearest neighbour interaction at fourth order. The three-loop graphs implement the constraint, as discussed in the text. }
\end{figure}

However, in general there are pathways contributing to the nn interaction which have no analog in the hopping pathways \footnote{At second order, no other interaction processes can occur. This is the reason for $\lambda=1$ at second order.}. These are processes in which \emph{none} of the dimers is static, and we have to calculate them diagrammatically. For nn, the corresponding diagrams are provided in Fig. \ref{TwoLoopAttractive}. We discuss the corresponding processes (the explicit calculations are performed in App. \ref{app:DSE}): In the left diagrams, the process starts with the decay of a dimer into atoms, where the sum over nearest neighbours indicates the $z=2d$ possibilities resulting from the unconstrained splitting vertex. The splitting is then followed by a double swap of atom and dimer. Finally, the atom on $k$ recombines with the one on the target site $i$ or $j$ into a dimer. The position indices of these processes are fixed by the nn range of the couplings, and no further summation occurs. Performing the frequency integrations for zero external frequencies, the left side of Fig. \ref{TwoLoopAttractive} evaluates to $-16 z Jx^3$, with $x = J/|U|$, thus providing an attraction piece to the dimer-dimer interaction. The three-loop diagrams on the left take care of the constraint: one decay possibility described by the unconstrained splitting vertex is not allowed, since the other dimer is located there. Indeed, the diagrams evaluate to $+16  Jx^3$, such that the net result of the processes in Fig. \ref{TwoLoopAttractive} is
\begin{eqnarray}
 \frac{v_a}{2} = - 16(z-1)Jx^3.
\end{eqnarray}
This is an attractive contribution to the nn interaction in any dimension. 

We are now in the position to provide the full fourth order contribution. The repulsive part to the interaction is given by the fourth order contribution to the nn hopping, which we can obtain from the expansion of Eq. \eqref{FullGreen} up to fourth order with the result
\begin{eqnarray}
\frac{v_r}{2}  = t = 2J[ x + (12 (z-1) -2) x^3] .
\end{eqnarray}
We thus obtain the final result for the nn hopping and interaction terms,
\begin{eqnarray}
t &=& 2J (x + 2(6 (z-1) -1)x^3) ,\\\nonumber
\frac{v}{2}&=& \frac{v_a + v_r}{2}  =2J (x + 2(2 (z-1) -1)x^3) ,\\\nonumber
\lambda &=& \frac{v}{2t}  =\frac{1 + 2(2 (z-1) -1)x^2}{1 + 2(6 (z-1) -1)x^2} \approx 1 - 8 (z-1) x^2 <1.
\end{eqnarray}

Finally, we discuss an additional effect at fourth order which is associated to the wave function renormalization. The leading term, relevant at fourth order perturbation theory, can be extracted from the frequency expansion of Eq. \eqref{ApproxGreen} at $\textbf{k}=0$, yielding
\begin{eqnarray}\label{WFR}
Z= 1 + 4 z x^2.
\end{eqnarray}
The correction to the dimensionless quantity $Z$ is $\mathcal O (x^2)$ and has effects at fourth order only. For example, if we are interested in the dispersion relation, obtained from the pole condition on the Green's function for real frequencies at a finite molecule momentum $\textbf{k}$, $G_d^{-1} (\mathrm i \omega(\textbf{k});\mu_b=E_b/2 ,\textbf{k})=0$, we find \footnote{The propagation of the dimers in vacuum is gapless, and we have adjusted the zero of energy at $\textbf{k}=0$ via appropriate choice of $\mu$.}
\begin{eqnarray}
 \omega  (\textbf{k}) =  \frac{4 J^2}{Z |U|} \sum_{\lambda} ( 1 - \cos \textbf{k} \textbf{e}_\lambda ).
\end{eqnarray}
The wave function contributes a fourth order term to the dispersion relation. In order to see the effect on the ratio of interaction vs. kinetic energy, we absorb the wave function renormalization factor into a redefinition of the field $t_2 \to \sqrt{Z} t_2$ and into the remaining couplings, e.g. $t \to \tilde t=t/Z, v\to \tilde v = v/Z^2$,  where the redefinition of the couplings is performed in such a way as to keep the full effective action invariant under the transformation. Since $Z>1$ we conclude that $\tilde \lambda = \tilde v/(2\tilde t) <\lambda$, such that the ratio of interaction vs. kinetic energy is effectively decreased additionally.  We finally note that the effects of the wave function renormalization correspond to an energy dependence of an effective Hamiltonian obtained in a Brillouin-Wigner perturbation theory for a microscopic Hamiltonian. It is well known that these effects occur at fourth order perturbation theory only.

\subsection{Relation of $n=0$ and $n=2$: Particle-Hole Mapping}
\label{sec:Di-Di}

At density $n=2$, the physics is expected to be similar to the case $n=0$. Similar to the latter, the effect of spontaneous symmetry breaking is absent due to the complete filling. Moreover, due to the constraint, there is only a single microscopic configuration, in full analogy to the completely empty lattice for $n=0$. Importantly, therefore the $n=2$ state in the constrained model must not be viewed as a Mott insulator phase, such as e.g. $n=1$ and strongly positive $U$, but rather as a constraint induced band insulator: For $n=1$ the motion of the particles simply costs energy (providing for the gap in the Mott spectrum), while for $n=2$ this is impossible in principle. Therefore, we may expect another ``vacuum problem''. The low lying excitations on top of this vacuum will be holes and doublets of holes or di-holes instead of atoms and dimers, and the complete amplitude resides in the $t_2$ mode. From this reasoning, we now replace 
\begin{eqnarray}
t_{2,i} \to 1 - t_{1,i}^\dag t_{1,i} - t_{0,i}^\dag t_{0,i} \equiv X'_i.
\end{eqnarray}
Alternatively, we could have adopted the formal point of view, applying the procedure described in Sec. \ref{sec:RotNew} for the angles $\chi= \theta =\pi$. Here we want to stress the physical similarities between $n=0,2$, but of course, the formal procedure generates exactly the same result. 

With the above replacement, the Hamiltonian takes the form
\begin{eqnarray}\label{HtHole}
H_{\text{kin}} &=& - J \sum_{\langle i,j\rangle}\big[ 2 t_{1,i}^\dag X'_i  X'_j t_{1,j}   +   t_{1,i}^\dag t_{0,i}t_{0,j}^\dag t_{1,j} \\\nonumber 
&& \qquad + \sqrt{2} ( X'_i  t_{1,i} t_{0,j}^\dag t_{1,j} +  t_{1,i}^\dag t_{0,i} t_{1,j}^\dag X'_j )  \Big], \\\nonumber
H_{\text{pot}} \hspace{-0.1cm}&=&\hspace{-0.1cm} +\mu \sum_i 2 t_{0,i}^\dag t_{0,i} + t_{1,i}^\dag t_{1,i}  - U\sum_{i} t_{0,i}^\dag t_{0,i} + t_{1,i}^\dag t_{1,i} \\\nonumber
&& + (- 2\mu + U) M^d,
\end{eqnarray}
where $M$ is the number of lattice sites in each lattice direction. The potential energy term can be written as 
\begin{eqnarray}
H_{\text{pot}} &=& -\mu' \sum_i 2 t_{0,i}^\dag t_{0,i} + t_{1,i}^\dag t_{1,i}  + U\sum_{i} t_{0,i}^\dag t_{0,i} ,\\\nonumber
\mu' &=& - \mu + U.
\end{eqnarray}
Working with the action in the following we may permute the operators at will. In this case, the mapping of the action for zero density to the one at $n=2$ is given, up to constants (the mean field energies) by the following replacements \begin{eqnarray}\label{replacements}
t_2&\to& t_0, \quad \mu \to \mu' , \quad U \to U, \quad g_{\text{split}}  \equiv \sqrt{2} J \to \sqrt{2}J, \nonumber\\
&& J_{\text{hop}} \equiv J \to 2 J,\quad  g_{\text{exchange}} \equiv 2J \to J.
\end{eqnarray}
While the coupling strength of the splitting vertex remains invariant under the transformation, the roles of $K^{(10)}$ and $K^{(21)}$ (cf. Eqs. (\ref{Ht},\ref{HC})) are exchanged, and so is the role of the corresponding coupling constants.
 
Based on Eq. \eqref{replacements} we observe that the scattering equation for the holes can be obtained from simple replacements in the scattering equations \eqref{ScattVac}. The equation for the hole splitting vertex,
\begin{eqnarray}\label{ScattVacHolesSplit}
\Gamma_\textbf{k}(\textbf{p})  &=& \Gamma^{(0)}_\textbf{k}(\textbf{p})  + \int\frac{d^dq}{(2\pi)^d} \frac{\Gamma_\textbf{k}(\bq) 2 (\Gamma^{(0)}_\textbf{k}(\textbf{p})  +\Gamma^{(0)}_\textbf{k}(\bq) )}{ E + 2\Gamma^{(0)}_\textbf{k}(\bq)  } \nonumber\\
\end{eqnarray} 
has the identical form to the one for dimers if we redefine the energy variable 
\begin{eqnarray}
E = \sqrt{2}( \mathrm i \omega - 2\mu ) \to E' =  \frac{\sqrt{2}}{2}\,\,(\mathrm i \omega - 2\mu' )
\end{eqnarray}
and therefore has the solution \eqref{FullYukawa} with the replacement $E\to E'$. The bare inverse di-hole propagator is $G_h^{(0) \,-1}(E',\textbf{k}) = \mathrm i \omega  + (-2\mu' + U) =  \mathrm i \omega  - (-2 \mu + U)$, and the di-hole self energy $\Sigma_h = G_h^{-1} - G_h^{(0) \,-1}$ is found to be
\begin{eqnarray}\label{ScattVacHolesSelf}
\Sigma_h(E',\textbf{k}) &=& 
   \frac 1 2\frac{1}{\sqrt{2}} ( I^{-1}(E') - E') \\\nonumber
&=&\frac{1}{4} \Big( - (\mathrm i \omega  -2 \mu'  ) \\\nonumber
&& + \Big[ \int\frac{d^dq}{(2\pi)^d} \frac{1}{- 4 (\epsilon_\bq  + \epsilon_{\bq - \textbf{k}}) + \mathrm i \omega - 2\mu'  } \Big]^{-1}\Big).
\end{eqnarray}
\emph{Bound state formation} -- The generalized ``Schr\"odinger equation'', governing the formation of a bound state of holes, is given by the pole condition
\begin{eqnarray}
G_h^{-1}(\omega = 0; \mu', \textbf{k} = 0) = 0 .
\end{eqnarray}
With definitions analogous to the zero density limit ($\mu' = \mu'_b - 2Jz, E'_b = 2\mu_b'$; this choice of $\mu_b'$ ensures the zero of kinetic energy to appear at zero momentum), and in dimensionless variables, we find the explicit form
\begin{eqnarray}\label{LSEHoles}
\frac{1}{ 4 | \tilde U| - 12 + 3\tilde{E}'_b} \hspace{-0.05cm}=\hspace{-0.1cm} \int\hspace{-0.1cm}\frac{d^dq}{(2\pi)^d} \frac{1}{ - \tilde{E}'_b + 4/d \sum_\lambda (1 - \cos \bq \textbf{e}_\lambda ) } .\nonumber\\
\end{eqnarray}
The bound state forms at the critical dimensionless interaction strength defined by $\tilde{E}'_b=0$. Due to the different structure of the scattering equations at $n=2$, we expect different results from the $n=0$ case. Indeed, we find the dimensionless critical interaction strengths in three and two dimensions
\begin{eqnarray} \label{onsets}
\tilde U'_3 \approx  - \frac{11}{3} \approx - 3.66 ,\quad \tilde U'_2 =-3 .
\end{eqnarray}
An interesting feature occurs in $d=2$: Despite the logarithmic infrared divergence of the fluctuation integral, the di-hole bound state formation occurs at a \emph{finite} attractive interaction strength. For the dependence of the binding energy on the interaction strength close to the threshold we find 
\begin{eqnarray}
d= 3 &:& \tilde{E}'_b = - 2\Big(\frac{2(|\tilde{U}|-|\tilde{U}'_3|)}{\sigma(|\tilde{U}'_3| -3)}\Big)^2,\\\nonumber 
d=2 &:& \tilde E'_b  \approx -\Lambda^2 \exp (- \tfrac{\pi}{ | \tilde U| - 3 }),
\end{eqnarray}
with $\sigma,\Lambda$ given below Eq. \eqref{QuadraticBinding}. In contrast, the mean field binding energy reads $\tilde E_b = - |\tilde U| + 4$.
The exact results Eq. \eqref{onsets} for the onset of the di-hole bound state can be compared to the mean field answer $\tilde U_\text{mf} = -4$. The shifts are substantially smaller than in the $n=0$ limit. Furthermore, from Fig. \ref{BindingEnergy} we observe that the nonperturbative fluctuation dominated quadratic and exponential regimes are smaller than in the low density limit described by \eqref{QuadraticBinding}. We conclude that fluctuations effects are weaker in the maximum density region $n\approx 2$ than in the low density counterpart $n=0$.

The finite value in Eq. \eqref{onsets} in two dimensions is surprising, and some comments are in order. An understanding is obtained from the fact that the coupling constants are different in the band insulator $n=2$ than in the vacuum $n=0$, and so the coefficients in the effective propagators are different. This, in particular, implies a shift in the effective interaction strength $|U| \to 4 |U| -  12 + 3\tilde E'_b$; note that the integrals are identical in Eq. \eqref{LSE3} and Eq. \eqref{LSEHoles} up to  a coefficient 4 instead of 2 in front of the kinetic term in the integral -- this reflects an enhanced mobility of the holes compared to the atoms. Therefore, we also have the log divergence in $d=2$ and $n=2$ for $\tilde E'_b \to 0$. However, this divergence now causes $|\tilde U| \to 3$ instead of $|U| \to 0$ for its compensation. We emphasize that it is not a matter of definition of the interaction strength: $U$ is a microscopic quantity, and the shift to it in Eq. \eqref{LSEHoles} is due to the effects of the $n=2$ band insulator. The finite threshold, thus, is an observable prediction of our theory. In addition, we would like to stress that there is no strong reason to expect identical critical interaction strengths at $n=0$ and $n=2$, as there is, in this regime of moderate interaction strengths, no particle-hole symmetry suggesting such behavior.

\emph{Effective di-hole theory} -- We have already noted that the mean field result for the binding energy is approached in the limit $U \to - \infty$. The leading correction is given by
\begin{eqnarray}
\Sigma_h(E'\to \infty ,\textbf{k}) \to  -  \frac{ 4J^2}{|U|} \sum_{\lambda} ( 1 + \cos \textbf{k} \textbf{e}_\lambda ) .
\end{eqnarray}
This precisely coincides with the leading perturbative self-energy correction for the dimers (cf. Eq. \eqref{ApproxGreen} at $\omega =0, -2\mu = |U|$). Furthermore, the di-hole-di-hole interaction is identical to the zero density case at $\mathcal O(J^2/|U|)$, since the splitting and the five-point vertices entering the dimer-dimer scattering vertex (cf. Fig. \ref{DimerDimer}) have the same value, and only the atomic onsite propagator enters at this order. Thus, to leading order we recover precisely the effective hardcore theory given by Eq. \eqref{SupplementAction} with $t_2$ (dimers) replaced by $t_0$ (di-holes) but identical interaction constants, though in general the original microscopic theories at $n=0,2$ feature different sets of interaction constants. The simple relation of the theory at $n=0$ and $n=2$ must be expected: When the atom-atom attraction $U$ is the largest scale in the problem, the integration of the high energy atom degrees of freedom in perturbation theory can be performed prior to the inclusion of the effects of, e.g. a finite density in the system. Due to the decoupling from the atomic degrees of freedom, the theories in the perturbative regime but at different densities must then be directly related to a reference density, e.g. $n=0$, by a rotation defined by Eq. \eqref{ratation}. The mapping of the theories for $n=0$ and $n=2$ may be understood as performing the rotation with $\theta =\chi =\pi$. This is different from the case where the atoms do not decouple completely, such that the density might have implicit fluctuation effects adding to the explicit effect of the rotation \eqref{ratation} on the dimer degrees of freedom. This hints at enhanced symmetry properties in the perturbative limit, which are discussed in the  companion paper \cite{Diehl09II}. 

Finally, we note that the low lying di-hole excitations on top of the $n=2$ band insulator state disperses quadratically, as expected for nonrelativistic excitations. Once the density is lowered away from $n=2$ and spontaneous symmetry breaking sets in, there  will be a linearly dispersing Goldstone mode \cite{Diehl09II}. 

\section{Conclusion}
\label{sec:Conclusion}

In this paper, we have developed a method which allows to exactly map a bosonic lattice model with three-body onsite constraint to a theory for two unconstrained bosonic degrees of freedom with conventional polynomial interactions. The Gutzwiller mean field theory is recovered as zero order contribution to the thermodynamic potential. The quadratic fluctuations around the mean field solution reproduce precisely spin wave theory. However, due to the exact nature of the mapping we can also systematically access the non-linearities. A convenient framework for our analysis is found to be the quantum effective action, the generating functional of the one-particle irreducible correlation functions. We establish that the usual symmetry principles are supplemented with a new ``constraint principle'', which depends on scale, being important at short distances but irrelevant at long wavelength. This setting allows to address fluctuations on various length scales in a unified framework. The application of the formalism to the full many-body problem is performed in the related work \cite{Diehl09II}, where we identify various quantitative and qualitative aspects which are uniquely tied to the presence of interactions. Here, in order to demonstrate the validity of the formalism, and to prepare for the many-body analysis in  \cite{Diehl09II}, we investigate the scattering properties of few particles in the limit of vanishing density. Furthermore, we address the complementary problem in  the limit of maximum filling, where the low lying excitations are holes and di-holes, and calculate the effective theory for hardcore dimers in the strong coupling limit.

We believe that the formalism developed here has strong potential applications beyond the particular model analyzed here and in the related Ref.  \cite{Diehl09II}. We conclude by giving an outlook on problems that may be addressed within this framework. 

\emph{Exactly constrained theories} -- Spin models with spin $S$. The onsite Hilbert spaces can always be formulated in terms of  $2S+1$ states. Here, we truncate to three states on each site ($S=1$), but a generalization is straightforward. Our construction maps such problems to $2S$ coupled bosonic degrees of freedom, and can thus be efficient for small total spin $S$. More specifically, our formalism can be particularly useful for $XXZ$ models in strong external fields and a perturbative anisotropy.
The mean field plus spin wave approximation is always incorporated in our setting, but it also offers the opportunity to systematically assess the interaction effects. We also stress that inhomogeneous ground states, such as e.g. a charge density wave  or antiferromagnet, are straightforwardly implemented in the formalism by applying site dependent rotations of the type \eqref{ratation}. Furthermore, it will be interesting to investigate if the limit of infinite spatial dimension $d\to \infty$ leads to a Gaussian fixed point of the theory, where the quadratic spin wave theory becomes exact. This would add the possibility of a $1/d$ expansion to the conventional $1/S$ (number of spin components) and $1/N$  (number of field components in the corresponding nonlinear sigma model) expansion used to treat spin models.

\emph{Models with an approximately realized constraint} --  A major challenge in quantum field theory is the treatment of strongly coupled systems where the non-linearity provides a larger energy scale than the scales appearing in the quadratic part.  
Our theoretical framework is suited to address a certain class of such strongly coupled systems, namely the one which is spanned by lattice models with strong local repulsion or attraction. It is not mandatory that the onsite repulsion (two-, three- or few- body) is infinitely large. Often, only a few dominant low energy degrees of freedom are necessary to capture the essential physics, and these degrees of freedom are explicitly implemented in our approach. For example, in the Mott phase of the Bose-Hubbard model, number fluctuations around the Mott state with given $n$ are small and a truncation to three Fock states $n,n\pm1$ is sensible \cite{Altman02}. Applying the formalism presented here, we are able to associate these particle and hole fluctuations with explicit bosonic degrees of freedom. In the resulting QFT, the nonperturbative calculation of the phase boundary in the repulsive Bose Hubbard model becomes feasible. Similarly, a strong but finite three-body repulsion $\gamma_3$ would lead to additional but massive bosonic degrees of freedom, which due to their mass $\sim \gamma_3$ can be taken into account perturbatively.

\emph{Fermions} -- The formalism is ideally suited to study attractive fermions on the lattice. In particular, recent experiments with 3-component fermions exhibit strong loss features \cite{Jochim08}  and are therefore important candidates for the observation of loss induced constrained models \cite{Kantian09}. A modified scheme for the implementation of the constraint in one dimension has already been given in  \cite{Kantian09}, followed by the analysis of the model with bosonization techniques. Our scheme, which explicitly introduces a mean field describing qualitative features of the ground state, is promising in higher spatial dimensions, and its improvements compared to Gutzwiller mean field theory may be expected to be similar to Dynamic Mean Field Theory \cite{KotliarDMFTReview}. We note that an application to the fermionic repulsive Hubbard model is complicated by the choice of the qualitative features of the ground state being debatable. In this case, a treatment of the constraint along the lines of  \cite{Muramatsu08} may be preferrable. 



\emph{Acknowledgements} -- We thank E. Altman, A. Auerbach, H. P. B\"uchler,  M. Fleischhauer, M. Greiter, A. Muramatsu, N. Lindner,  J. M. Pawlowski, L. Radzihovsky, S. Sachdev, J. Taylor and C. Wetterich for interesting discussions. This work was supported by the Austrian Science Foundation through SFB F40 FOQUS, by the European union via the  integrated project SCALA, by the Russian Foundation for Basic Research, and by the Army Research Office with funding from the DARPA OLE program.

\begin{appendix}

\section{Dyson-Schwinger Equations}
\label{app:DSE}

We calculate the DSE for the two- and three-point functions exactly, and the four-point function perturbatively. In vacuum, a tremendeous simplification is provided by the fact that density-type contributions vanish. In the loop language, this translates to checking whether there is a subdiagram in which the arrows form a closed tour. 

\emph{Formalism} -- The most important step is to find the correct symmetry factors for the loop contributions. For that purpose, it is sufficient to consider the field index $\alpha$ to parameterize the field type only; the frequency and momentum structure of the couplings can then be found from the corresponding conservation laws in a second step, reading off the couplings from the microscopic action \eqref{FourierAct}. Thus we consider
\begin{eqnarray}
\chi_\alpha = (t_1, t_1^\dag, t_2 , t_2^\dag) .
\end{eqnarray}
In this formalism, we find the following couplings:
\begin{eqnarray}
S^{(2)}_{\alpha \beta} \equiv \frac{\delta^2 S }{\delta t_\alpha\delta t_\beta}  &=& p_1 [\delta_{\alpha 1}\delta_{\beta 2} + {\rm perm.}] \\\nonumber
&& +  p_2 [\delta_{\alpha 3}\delta_{\beta 4} + {\rm perm.}] ,
\end{eqnarray}
where ''perm." means to permute all greek indices while leaving the explicit numbers in their positions. This coupling must be inverted to give the propagators, with the result
\begin{eqnarray}
G_{\alpha \beta }  =  \frac{1}{p_1} [\delta_{\alpha 1}\delta_{\beta 1} + {\rm perm.}] + \frac{1}{p_2} [\delta_{\alpha 3}\delta_{\beta 4} + {\rm perm.}] .
\end{eqnarray}
The higher couplings read
\begin{eqnarray}
S^{(3)}_{\alpha \beta\gamma} \hspace{-0.2cm}&=&\hspace{-0.1cm} [\delta_{\alpha 1}\delta_{\beta 1} \delta_{\gamma 4} + \delta_{\alpha 2}\delta_{\beta 2} \delta_{\gamma 3} + {\rm perm.}],\\\nonumber
S^{(5)}_{\alpha \beta\gamma\delta\epsilon} \hspace{-0.2cm}&=&\hspace{-0.1cm}  s^{(5)}[\delta_{\alpha 1}\delta_{\beta 1} \delta_{\gamma 4} \delta_{\delta 4}\delta_{\epsilon 3}  +\delta_{\alpha 2}\delta_{\beta 2} \delta_{\gamma 3} \delta_{\delta 4}\delta_{\epsilon 3}  + {\rm perm.}].
\end{eqnarray}
$p_1,p_2,s^{(3)},s^{(5)}$ can be read off directly from the classical action \eqref{FourierAct} after proper symmetrization in momentum space. We can reduce the tensors by contracting with delta functions. For example, 
\begin{eqnarray}
S^{(3)}_{1 1 4} &=& s^{(3)}\delta_{\alpha 1}\delta_{\beta 1} \delta_{\gamma 4}S^{(3)}_{\alpha \beta \gamma} =s^{(3)} \delta_{\alpha 1}\delta_{\beta 1}[ \delta_{\alpha 1}\delta_{\beta 1} +  {\rm perm.}] \nonumber\\
&=& 2 s^{(3)}\delta_{\alpha 1} \delta_{\alpha 1} = 2s^{(3)} .
\end{eqnarray}
where doubly occuring indices are summed over. The couplings are now fully symmetrized in field space, and indices may be permuted at will.

\emph{Two- and three-point functions} -- We consider the DSEs for the two- and the three point functions for the inverse propagators and the splitting vertex. In the vacuum limit, these equations are one-loop:
\begin{eqnarray}
\Gamma^{(2)}_{\alpha\beta} &=& S^{(2)}_{\alpha\beta} - \frac{1}{2} \mathrm{tr} \, S^{(3)}_{\alpha \mu_1\mu_2} G_{\mu_1\nu_1} G_{\nu_2\mu_2}\Gamma^{(3)}_{\beta \nu_1\nu_2} ,\\\nonumber
\Gamma^{(2)}_{12} \equiv p_1 &=& p_{1,0}  ,\,\,
\Gamma^{(2)}_{34} \equiv p_1  = p_{2,0} - 2 \mathrm{tr} \,s^{(3)} \frac{1}{p_1} \frac{1}{p_1}\gamma^{(3)} ,\\\nonumber
\Gamma^{(3)}_{114} &=& S^{(3)}_{114} - \frac{1}{2}\mathrm{tr} \, S^{(4)}_{11\mu_1\mu_2} G_{\mu_1\nu_1} G_{\nu_2\mu_2}\Gamma^{(3)}_{4 \nu_1\nu_2}  \\\nonumber
&=& 2\gamma^{(3)}  = 2 s^{(3)}  - 4 \mathrm{tr}\, s^{(4)} \frac{1}{p_1} \frac{1}{p_1}\gamma^{(3)}.
\end{eqnarray}
tr indicates the frequency and momentum integrations, while the discrete index contractions are carried out explicitly. The corresponding integral equations for specific external momentum configurations relevant for two-body scattering are given in Eq. \eqref{ScattVac} resp. Fig. \ref{fig1}. $\Gamma^{(3)}_{114}  = 2\gamma^{(3)}$ holds under the assumption that the full Feshbach vertex has the same structure in field space as the classical one. The atom propagator is not renormalized in vacuum. Note that only the atom propagator enters the diagrams for the dimer self energy and the splitting vertex.

\emph{Four-point functions} -- We further want to compute the fourth order interaction vertices.\\
1) Second order, momentum space calculation --  If we restrict to the order $J^2$, we can limit ourselves to two-loop order. We find
\begin{widetext}
\begin{eqnarray}\label{GammaFour}
\Gamma^{(4)}_{\alpha \beta\gamma\delta} &=& S^{(4)}_{\alpha \beta\gamma\delta} - \frac{1}{2}\mathrm{tr} \, \big[ S^{(5)}_{\alpha \beta\gamma\mu_1\mu_2} G_{\mu_1\nu_1} G_{\nu_2\mu_2}\Gamma^{(3)}_{\delta \nu_1\nu_2}  
+  S^{(5)}_{\alpha \beta\delta\mu_1\mu_2} G_{\mu_1\nu_1} G_{\nu_2\mu_2}\Gamma^{(3)}_{\gamma \nu_1\nu_2}\\\nonumber
&&\qquad\qquad + S^{(5)}_{\alpha \gamma\delta\mu_1\mu_2} G_{\mu_1\nu_1} G_{\nu_2\mu_2}\Gamma^{(3)}_{\beta \nu_1\nu_2}
+ \Gamma^{(5)}_{ \beta\gamma\delta\mu_1\mu_2} G_{\mu_1\nu_1} G_{\nu_2\mu_2}\Gamma^{(3)}_{\beta \nu_1\nu_2}\big] \\\nonumber
&&- \frac{1}{3!}\mathrm{tr} \, \big[  S^{(5)}_{ \beta\gamma\mu_1\mu_2\mu_3} G_{\mu_1\nu_1} G_{\nu_2\mu_2}G_{\mu_3\nu_3}\Gamma^{(5)}_{\alpha\delta \nu_1\nu_2\nu_3} 
+ S^{(5)}_{ \beta\delta\mu_1\mu_2\mu_3} G_{\mu_1\nu_1} G_{\nu_2\mu_2}G_{\mu_3\nu_3}\Gamma^{(5)}_{\alpha\gamma \nu_1\nu_2\nu_3} \\\nonumber
&&\qquad + S^{(5)}_{ \alpha\beta\mu_1\mu_2\mu_3} G_{\mu_1\nu_1} G_{\nu_2\mu_2}G_{\mu_3\nu_3}\Gamma^{(5)}_{\gamma\delta \nu_1\nu_2\nu_3}\big].
\end{eqnarray}
\end{widetext}
Working at second order the full vertices are replaced by the classical ones. The dimer-dimer interaction then evaluates to
\begin{eqnarray}
\Gamma^{(4)}_{4343} = 4 \gamma^{(4)}  &=& - 8 \mathrm{tr} \, [ s^{(5)} \frac{1}{p_1} \frac{1}{p_1}s^{(3)}  +  s^{(3)} \frac{1}{p_1} \frac{1}{p_1}s^{(5)}] \nonumber \\
&&- 8  \mathrm{tr} \, s^{(5)} \frac{1}{p_1} \frac{1}{p_{2,0}}\frac{1}{p_1}s^{(5)} .
\end{eqnarray}
The first two contributions are one-loop, the last one is two-loop. 

Now we insert the appropriate momentum structures. At order $J^2$ the $J$ dependence of the atom propagator is to be neglected and the dimer propagator equals the bare one, 
\begin{eqnarray}
p_1(\omega ) &=& \mathrm i \omega - \mu = \mathrm i \omega +|U|/2,\\\nonumber
p_{2}(\omega )&=& p_{2,0}(\omega )= \mathrm i \omega -2\mu +U = \mathrm i \omega .
\end{eqnarray}
We then find, for zero external frequency,
\begin{widetext}
\begin{eqnarray}
\Gamma^{(4)}_{4343}  &=&+16 J^2 \big( \int_q \frac{1}{p_1( \omega)}  \frac{1}{p_1( -\omega)} 
\big[\tfrac{1}{2} [\epsilon_{\bq_3 - \bq} + \epsilon_{\bq + \bq_3 - \bq_4} + \epsilon_{\bq_1 - \bq} + \epsilon_{\bq + \bq_1 - \bq_4}][\epsilon_{ \bq - \bq_4 } + \epsilon_{\bq}] + \{\bq_3 \leftrightarrow \bq_4,\bq_1 \rightarrow \bq_2\}\big]\nonumber\\\nonumber
&&  \qquad - \int_{q,q'} \frac{1}{p_1( \omega)}  \frac{1}{p_1( \omega')}  \frac{1}{p_2(- \omega - \omega') } \tfrac{1}{2} [\epsilon_{\bq_2 - \bq} + \epsilon_{\bq_2 - \bq' } + \{\bq_2  \rightarrow \bq_4\}]
\tfrac{1}{2} [\epsilon_{\bq_1 - \bq} + \epsilon_{\bq_1 - \bq' } + \{\bq_1  \rightarrow \bq_3\}]\big)\\\nonumber
&=& 8 \frac{J^2}{|U|} [\epsilon_{\bq_1}+ \epsilon_{\bq_2} + \epsilon_{\bq_3} + \epsilon_{\bq_4} + \epsilon_{\bq_1 - \bq_2} +\epsilon_{\bq_1 - \bq_4} ] = 4\gamma^{(4)}.
\end{eqnarray}
The contribution to the effective action then reads 
\begin{eqnarray}
\Delta S &=&\frac{2J^2}{|U|}  \int\limits_{q_1, ... ,q_4}\delta ( q_1 - q_2 + q_3 - q_4)) t^\dag_{2 }(q_1)t_{2 }(q_2)t^\dag_{2}(q_3)t_{2}(q_4) [\epsilon_{\bq_1}+ \epsilon_{\bq_2} + \epsilon_{\bq_3} + \epsilon_{\bq_4} + \epsilon_{\bq_1 - \bq_2} + \epsilon_{\bq_1 - \bq_4} ] \\\nonumber
&=& \frac{2J^2}{|U|}\sum_{\langle i,j\rangle} \big[ t_{2,i}^\dag (\tau)  t_{2,j}(\tau)\big[  n_{2,i}(\tau) + n_{2,j}(\tau)] + n_{2,i}(\tau)n_{2,j}(\tau)\big].
 \end{eqnarray}
 \end{widetext}

2) Fourth order, real space nearest neighbours -- Finally, we compute the diagrams displayed in Fig. \ref{TwoLoopAttractive}. As discussed in the text, this is not the full contribution to the interaction vertex, which has a high diagrammatic complexity. (It could be obtained from Eq. \eqref{GammaFour}, to fourth order, but requires knowledge of $\Gamma^{(5)}$ to second order.) However, knowing which diagrams we want to compute the procedure outlined above allows to identify the correct symmetry factors and contraction prescriptions. Since we are interested in a process on nn sites $i,j$, it is more efficient to do the calculation in frequency and \emph{real} space, fixing the external legs at the nearest neighbour configuration, cf. Fig. \ref{TwoLoopAttractive}. We find
\begin{widetext}
\begin{eqnarray}
\gamma^{(4)} |_{ij;ij}&\ni&   - 2 \mathrm{tr} \, [ s^{(3)} \frac{1}{p_1} s^{(4)} \frac{1}{p_1} \frac{1}{p_2}s^{(4)}\frac{1}{p_1}s^{(3)} \frac{1}{p_1} ]
+2 \mathrm{tr} \, [ s^{(5)} \frac{1}{p_1}  \frac{1}{p_2} s^{(4)}  \frac{1}{p_1}  \frac{1}{p_2} s^{(4)} \frac{1}{p_2} s^{(4)} \frac{1}{p_1} s^{(3)} \frac{1}{p_1} ] \\\nonumber
&=& - 16 J^4 z \int\frac{d\omega }{2\pi}  \int\frac{d\omega' }{2\pi} \frac{1}{p_1( \omega  )} \frac{1}{p^2_1( -\omega  )}\frac{1}{p_1( \omega'  )}\frac{1}{p_2(- \omega - \omega'  )} \\\nonumber
&& +  16 J^4  \int\frac{d\omega }{2\pi}  \int\frac{d\omega' }{2\pi} \int\frac{d\omega'' }{2\pi} \frac{1}{p_1( -\omega - \omega'  )} \frac{1}{p_1( \omega''   )}\frac{1}{p_2(\omega +  \omega' - \omega''  )}\frac{1}{p_1( \omega  )} \frac{1}{p_2(  \omega'  )}\frac{1}{p_1( \omega + \omega'  )}\\\nonumber
&=& -16 (z - 1) \frac{J^4}{|U|^3},
\end{eqnarray}
\end{widetext}
where the full vertices have been replaced by the bare counterparts and $p_1(\omega ) = \mathrm i \omega - \mu = \mathrm i \omega +|U|/2, p_{2}(\omega )= \mathrm i \omega$, as appropriate at this order perturbation theory. The overall factor of two in the first line may be interpreted in terms of the two rows in Fig. \ref{TwoLoopAttractive}. The factor of $z$ in two-loop diagram results from the $z$ possibilities for a dimer to decay via the bare splitting vertex. No further summations occur in real space. The three-loop diagram corrects for the fact that one of these possibilities is not allowed due to the presence of the other dimer. 

\end{appendix}

\end{document}